\documentclass[reprint,twocolumn,amsmath,amssymb,aps,nofootinbib,longbibliography]{revtex4-1}
\pdfoutput=1
\usepackage{tablefootnote}
\usepackage[T1]{fontenc}
\usepackage{adjustbox}
\usepackage{multirow}
\usepackage{hyperref}
\usepackage{bbm}
\usepackage{xcolor}
\newcommand{\yong}[1]{{\color{black}#1}}
\newcommand{\eqal}[1]{\begin{align}#1\end{align}}
\newcommand{\neff}{$N_{\rm eff}$ }
\newcommand{\neffe}{$N_{\rm eff}$}
\newcommand{\nb}{\nonumber}
\newcommand{\ket}{\rangle}
\newcommand{\bra}{\langle}

\begin{document}

\title{$N_{\rm eff}$ as a new physics probe in the precision era of cosmology}
\author{Yong Du}
\affiliation{Tsung-Dao Lee Institute \& School of Physics and Astronomy, Shanghai Jiao Tong University, China}
\email{yongdu5@sjtu.edu.cn}

\date{\today}
             
\begin{abstract}
We perform a global fit to the electroweak vertices and 4-fermion operators of the standard model effective field theory in this work using \neff from cosmological probes, as well as data sets from colliders and low-energy experiments.
We find \neffe, both its current measurement and future projections, can only marginally improve the fit in both the flavor universal and the most general flavor scenarios.
The resulting $1\sigma$ bound on \neff is significantly improved from the global fit and becomes comparable to its current theoretical uncertainty, such that the latter will become important for this study at next generation experiments like future lepton colliders.
\neff from the global fit is also adopted to predict the primordial helium abundance $Y_P$, which significantly reduces the parameter space on the $Y_P$-\neff plane.
Through error propagation, we also conclude that reducing the experimental uncertainty of $Y_P$ from metal-poor galaxies down below 0.2\% could play an important role in deepening our understanding on the free neutron lifetime anomaly.
\end{abstract}

\maketitle

\section{Introduction}\label{sec:intro}
The standard model (SM) of particle physics has been very successful in explaining almost all experimental data up to now, yet it is also broadly accepted that the SM cannot be the complete theory as it does not provide any explanation for several puzzles: the nature of dark matter, the masses of neutrinos, the origin of nucleon mass and spin, the quantum nature of gravity {\textit{etc}}.

While many well-motivated new models can be constructed to explain the unresolved aforementioned puzzles, it is unclear which one would be realized in nature due to the null observation of any new particles at the LHC since the discovery of the Higgs boson in 2012\,\cite{ATLAS:2012yve,CMS:2012qbp}. With its increasing luminosity and center-of-mass energy in particular, the LHC is gradually pushing the new physics scale above around 1\,TeV, a scale much larger than the weak one. This, in turn, suggests that at the weak scale, we can integrate the underlying new heavy physics out and work in its low-energy effective theory that contains the standard matter contents of the SM only. The apparent difference with the SM is that we now have contact interactions with mass dimension larger than 4. Generically, we can parameterize the corresponding Lagrangian of this kind of effective field theories as
\eqal{
\mathcal{L} = \mathcal{L}_{\rm SM} + \sum_{n=5}C_{4-n} \mathcal{O}_n,
}
with $n$ the mass dimension of the operator $\mathcal{O}_n$, and $(4-n)$ that of the Wilson coefficient $C_{4-n}$. In this work, we will work in the SM Effective Field Theory (SMEFT) that respects the same local gauge group of the SM with the same field components, and only consider the leading order corrections from these operators. Recall that the dimension-5 operators only contribute to neutrino masses\,\cite{Weinberg:1979sa}, this approximation thus means that we will only consider dimension-6 SMEFT operators\,\cite{Buchmuller:1985jz,Grzadkowski:2010es} in this work.

These dimension-6 operators will, on the one hand, change the SM interaction strength, and on the other hand, introduce contact interactions that are absent in the SM Lagrangian. One well-known example is the semi-leptonic 4-fermion operator $\bar ude\bar\nu$ responsible for beta decay as proposed by Fermi in 1933\,\cite{Fermi:1934sk}. In a similar pattern and in the SMEFT framework, pure leptonic 4-fermion operators like $\bar l l \bar l l$ will be present. These 4-fermion operators, together with those modifying the SM vertices, could affect lepton pair production measured at LEP\,\cite{ALEPH:2013dgf} for instance. Provided these corrections from the SMEFT operators are large, these deviations from the SM prediction will have the chance to be observed experimentally. Alternatively, if no significant deviation is captured by detectors, constraints can be put on the corresponding Wilson coefficients.

Due to the high energy and/or the high luminosity, tremendous studies utilizing the SMEFT framework have been carried out given the very rich data from LEP and/or the LHC, see for example \cite{Efrati:2015eaa,Falkowski:2015krw,Berthier:2016tkq,Falkowski:2017pss,Barklow:2017suo,Barklow:2017awn,Cirigliano:2018dyk,DeBlas:2019qco,Ellis:2020unq,Gu:2020ldn,Ethier:2021bye,deSalas:2021aeh,Cirigliano:2021yto,Durieux:2022cvf,Ellis:2022zdw,Aoude:2022deh,Schwienhorst:2022yqu,Breso-Pla:2023tnz,Ellis:2023ucy} and references therein. As mentioned earlier, since the SM has been working so well to fit these data, results are generically presented as constraints on the Wilson coefficients. While it is a common practice in literature to consider one operator at a time in this procedure, a model-independent study shall include all contributing operators since in the underlying model(s), different operators are likely interdependent. This in turn motivates us to perform a global fit to the SMEFT operators when doing such an exercise.

In this work, we will combine colliders with cosmological probes and consider SMEFT global fit during neutrino decoupling in the early Universe. At the beginning, neutrinos were in thermal equilibrium with the other components in the thermal plasma through weak interactions. However, as the Universe expands, the temperature drops such that when the temperature is below about 2\,MeV\,\cite{Dolgov:2002wy}, this equilibrium could not be maintained any longer and neutrinos then decoupled from the rest of the plasma and started free streaming. In the SMEFT, several types of new interactions between neutrinos, electrons, as well as neutrinos and electrons, are present. As a result, the evolution history of neutrinos in the early Universe will be modified and neutrinos could decouple from the plasma at a different epoch than that in the standard scenario. Consequently, \neffe, the effective number of relativistic species in the early Universe, will be predicted differently, and this difference may then be detected from precision cosmology. Theoretically, \neff has recently been calculated with a precision of $\mathcal{O}(10^{-4})$\,\cite{Akita:2020szl,Froustey:2020mcq,Cielo:2023bqp}. Experimentally, though the current Planck uncertainty is still at the 10\% level, the planned next generation experiments such as SPT-3G\,\cite{Benson:2014qhw}, CMB-S4\,\cite{Abazajian:2016yjj}, CORE\,\cite{DiValentino:2016foa}, the Simon Observatory\,\cite{Ade:2018sbj}, PICO\,\cite{Hanany:2019lle} and CMB-HD\,\cite{Sehgal:2019ewc,CMB-HD:2022bsz} would be able to eventually reduce the uncertainty down below the percent level. This therefore motivates us to perform the global fit of the SMEFT mentioned at the beginning of this paragraph.

By including \neff in the global fit, we then find that
\begin{itemize}
\item using the numerical expression in eq.\,\eqref{eq:neffsmeft}, the inclusion of \neff only marginally improves the SMEFT global fit in both the flavor universal and the most general flavor scenarios;
\item the $1\sigma$ bound on \neff from the global fit is reduced significantly and becomes comparable to its current theoretical uncertainty. \yong{Consequently, the theoretical uncertainty of \neff can no longer be ignored when taking into account the data sensitivity for next generation experiments like future lepton colliders.} In addition, \neff from the global fit also significantly narrows down the parameter space on the $Y_P$-\neff plane, as shown by the inset in the upper right corner of Fig.\,\ref{fig:neff2yp}.
\item improving the extraction precision of the primordial abundance of helium down below 0.2\% experimentally could help answer if new ingredients beyond the standard big-bang nucleosynthesis are needed in understanding the free neutron lifetime anomaly, as illustrated in Fig.\,\ref{fig:neff2yp}.
\end{itemize}

In obtaining these results, we firstly discuss \neff calculation within the SM and in the presence of SMEFT operators in section\,\ref{sec:neffsm}, we discuss the input parameters used in this study in the end, and then show our results in section\,\ref{sec:results}. In presenting our results, we show the numbers in both the flavor universal scenario and the flavor general scenario. Then in section\,\ref{sec:ypneutron}, we discuss how the global fit could affect the prediction of the primordial abundance of helium, as well as its connection to the free neutron lifetime anomaly. We then conclude in section\,\ref{sec:conclu}.

\section{\boldmath $N_{\rm eff}$ in the SMEFT}\label{sec:neffsm}
In this section, we will firstly briefly review the precision calculation of \neff in the SM, and then introduce corrections from the dimension-6 operators in the SMEFT. Given our treatment on lepton flavors, a brief comment on $\nu_{\mu,\tau}$ will also be discussed.

\subsection{\boldmath Brief review of $N_{\rm eff}$ in the SM}\label{}
Neutrinos, electrons and positrons, as well as photons are in thermal equilibrium in the very early Universe with an equal temperature, i.e., $T_\gamma = T_e = T_\nu$. However, as the Universe expands, weak interactions start to decouple from the thermal plasma at around $T_{\rm dec}=2$\,MeV\,\cite{Dolgov:2002wy} when the weak interaction strength falls below the Hubble parameter. At the same time, photons and electrons are still tightly coupled with each other with $T_\gamma = T_e$ thanks to quantum electrodynamics (QED). The neutrinos will simply undergo dilution after the decoupling, while the photons can still receive energy injection from the plasma even when the temperature of the Universe drops below $m_e$ from $e^+ e^-\to \gamma \gamma$.\footnote{Note that when the temperature of the Universe is below $m_e$, the inverse process $\gamma \gamma \to e^+ e^-$ will be suppressed by $e^{-m_e/T}$.} As a result, the photon temperature will be slightly higher than that of the neutrino, which can be parameterized as\,\cite{Shvartsman:1969mm,Steigman:1977kc,Mangano:2001iu}:
\eqal{
\rho_R=\left[ 1+\frac{7}{8}\left(\frac{4}{11}\right)^{\frac{4}{3}} N_{\rm eff}\right]\rho_\gamma.
}
Here, $\rho_\gamma=(\pi^2/30)g_\gamma T_\gamma^4$ is the photon energy density, $\rho_R = \rho_\gamma + 3\rho_\nu + 3\rho_{\bar\nu}$ with $\rho_\nu = \rho_{\bar\nu} =(7\pi^2/240)g_\nu T_\nu^4$ is the total energy density of all relativistic species at the considered time,\footnote{We ignore neutrino masses due to their tininess.} whose effective number is denoted as \neffe, and $g_i$ is the intrinsic degrees of freedom of particle $i$. In the instantaneous neutrino decoupling limit, it is well-known that $N_{\rm eff}=3$, while in general, it is straightforward to solve from the above equations that
\eqal{
N_{\rm eff}=3\times\left(\frac{11}{4}\right)^{\frac43} \times \left(\frac{T_{\nu}}{T_{\gamma}}\right)^{4}.\label{eq:nefffinal}
}
Taking the neutrino non-instantaneous decoupling effects, neutrino oscillations during neutrino decoupling\,\cite{Hannestad:2001iy,Dolgov:2002ab,Mangano:2005cc,deSalas:2016ztq,Gariazzo:2019gyi}, as well as finite temperature QED corrections\,\cite{Heckler:1994tv,Fornengo:1997wa,Bennett:2019ewm} into account, it has been calculated that $N_{\rm eff}=3.044$\,\cite{Akita:2020szl,Froustey:2020mcq}, with higher-order theoretical corrections at $\mathcal{O}(10^{-4})$ and can be safely neglected for this study.

Therefore, at this stage, it is clear that to precisely predict \neff at any time during the evolution of the Universe, one shall track the evolution of $T_\gamma$ and $T_\nu$ and then make use of eq.\,\eqref{eq:nefffinal}. On the other hand, the evolution of $T_\gamma$ and $T_\nu$ can be obtained by solving the Boltzmann equations, which read
\eqal{
\frac{d n}{d t}+3 H n &=\frac{\delta n}{\delta t}\equiv\int g\frac{d^{3} p}{(2 \pi)^{3}} \mathcal{C}[f],\label{numdensity} \\
\frac{d \rho}{d t}+3 H(\rho+p) &=\frac{\delta \rho}{\delta t}\equiv\int g E \frac{d^{3} p}{(2 \pi)^{3}} \mathcal{C}[f],\label{energydensity}
}
where $g$ is the intrinsic degrees of freedom of the particle under consideration, $E$ is its energy, and $n$ and $\rho$ are the number and the energy densities, respectively. The collision term $\mathcal{C}$ is given by
\begin{widetext}
\eqal{
\mathcal{C}\left[f_{i}\right] \equiv & \frac{1}{2 E_{i}} \sum_{X, Y} \int \prod_{i,j} d \Pi_{X_{i}} d \Pi_{Y_{j}}(2 \pi)^{4} \delta^{4}\left(p_{i}+p_{X}-p_{Y}\right) \nonumber\\
& \times\left( \langle\mathcal{M}^{2}\rangle_{Y \rightarrow i+X} \prod_{i,j} f_{Y_{j}}\left[1 \pm f_{i}\right] \left[1 \pm f_{X_{i}}\right] -  \langle\mathcal{M}^{2}\rangle_{i+X \rightarrow Y}  \prod_{i,j} f_{i} f_{X_{i}}\left[1 \pm f_{Y_{j}}\right]\right),\label{collterm}
}
\end{widetext}
with $d \Pi_{{i}}\equiv d^3p_{i}/[(2\pi)^3 2E_{i}]$, and ``+ ($-$)'' for bosonic (fermionic) particles. The evolution of $T_\gamma$, $T_\nu$, and the chemical potential $\mu$ can then be obtained through applying chain rules to eq.\,(\ref{numdensity}-\ref{energydensity}) to give\,\cite{Escudero:2020dfa}\footnote{Note that the photon chemical potential $\mu_\gamma=0$ and that of the electron can be safely set to zero for the reason that $\mu_e\ll T_\gamma$. Thus, in practice, we only need to numerically solve the neutrino chemical potentials, whose impact on \neff turns out small as we will see below, and the temperatures by assuming thermal equilibrium phase space distribution for each species.}
\begin{widetext}
\eqal{
\frac{d T_{\gamma}}{d t} = &\, -\frac{4 H \rho_{\gamma}+3 H\left(\rho_{e}+p_{e}\right)+3 H T_{\gamma} \frac{d P_{\text {int }}}{d T_{\gamma}}+\frac{\delta \rho_{\nu e}}{\delta t} + \frac{\delta \rho_{\nu \mu}}{\delta t} + \frac{\delta \rho_{\nu \tau}}{\delta t}}{\frac{\partial \rho_{\gamma}}{\partial T_{\gamma}}+\frac{\partial \rho_{e}}{\partial T_{\gamma}}+T_{\gamma} \frac{d^{2} P_{\text {int }}}{d T_{\gamma}^{2}}},\label{TgammaFinal}\\
\frac{d T_{\nu_{\alpha}}}{d t} = &\, -H T_{\nu_{\alpha}}+\frac{\delta \rho_{\nu_{\alpha}}}{\delta t} / \frac{\partial \rho_{\nu_{\alpha}}}{\partial T_{\nu_{\alpha}}}, \quad \quad \alpha = e,\mu,\tau\label{TvFinal},\\
\frac{d \mu_{\nu_\alpha}}{d t} = &\, \frac{-1}{\frac{\partial n_{\nu_\alpha}}{\partial \mu_{\nu_\alpha}} \frac{\partial \rho_{\nu_\alpha}}{\partial T_{\nu_\alpha}}-\frac{\partial n_{\nu_\alpha}}{\partial T_{\nu_\alpha}} \frac{\partial \rho_{\nu_\alpha}}{\partial \mu_{\nu_\alpha}}}\left[-3 H\left((p_{\nu_\alpha}+\rho_{\nu_\alpha}) \frac{\partial n_{\nu_\alpha}}{\partial T_{\nu_\alpha}}-n_{\nu_\alpha} \frac{\partial \rho_{\nu_\alpha}}{\partial T_{\nu_\alpha}}\right) + \frac{\partial n_{\nu_\alpha}}{\partial T_{\nu_\alpha}} \frac{\delta \rho_{\nu_\alpha}}{\delta t}-\frac{\partial \rho_{\nu_\alpha}}{\partial T_{\nu_\alpha}} \frac{\delta n_{\nu_\alpha}}{\delta t}\right] \label{chemevolv}.
}
\end{widetext}
Here, $P_{\rm int}$ and $\rho_{\rm int}\equiv -P_{\rm int} + dP_{\rm int}/d\ln T_\gamma$, calculated in \cite{Bennett:2019ewm}, are finite temperature corrections to the electromagnetic pressure and energy density, respectively. Solving eqs.\,(\ref{TgammaFinal}-\ref{chemevolv}) is in general quite challenging and time-consuming since, for any $2\to2$ processes, the collision term integrals are 12-fold ones whose analytical expressions are not known. In the EFT language as we adopt in this work, a complete, analytical, and generic dictionary for the collision term integrals have been built in \cite{Du:2021idh} up to dimension 7, rendering numerically solving eqs.\,(\ref{TgammaFinal} -\ref{chemevolv}) or \neff efficient when including dimension-6 operators in the SMEFT.
\begin{table}[!htb]\caption{$\nu_e$ and $\nu_{\mu,\tau}$ scattering/annihilating processes contributing to non-instantaneous neutrino decoupling in the early Universe. For each process above, the incoming neutrino can be replaced by its anti-partner, which also contributes to \neff and is implicitly implied in the table above.}\label{tab:procs}
\centering{
\begin{tabular}{c|c|c}
\hline\hline
\multicolumn{3}{c}{Scattering/Annihilation Processes}\\
\hline
Label & $\nu_e$ case ($j={\mu,\tau}$) & $\nu_i$ case ($i={\mu,\tau}$; $j=i\cup{e}$, $i\ne j$)\\ 
\hline 
1 & $ \nu_e + \bar\nu_e \to \nu_e + \bar\nu_e$  &  $ \nu_i + \bar\nu_i \to \nu_i + \bar\nu_i$ \\ 
2 & $ \nu_e + \nu_e \to \nu_e + \nu_e$  &  $ \nu_i + \nu_i \to \nu_i + \nu_i$ \\   
3 & $ \nu_e + \bar\nu_e \to \nu_j + \bar\nu_j$  &  $ \nu_i + \bar\nu_i \to \nu_j + \bar\nu_j$ \\ 
4 & $ \nu_e + \bar\nu_j \to \nu_e + \bar\nu_j$  &  $ \nu_i + \bar\nu_j \to \nu_i + \bar\nu_j$ \\ 
5 & $ \nu_e + \nu_j \to \nu_e + \nu_j$  &  $ \nu_i + \nu_j \to \nu_i + \nu_j$ \\ 
6 & $ \nu_e + \bar\nu_e \to e^+ + e^-$  &  $ \nu_i + \bar\nu_i \to e^+ + e^-$ \\ 
7 & $ \nu_e + e^- \to \nu_e + e^-$  &  $ \nu_i + e^- \to \nu_i + e^-$ \\ 
8 & $ \nu_e + e^+ \to \nu_e + e^+$  &  $ \nu_i + e^+ \to \nu_i + e^+$ \\ 
\hline\hline
\end{tabular}
}
\end{table}

In the SM, the collision term integrals come from the scattering/annihilation processes summarized in Table\,\ref{tab:procs}. All these processes are mediated by an off-shell $Z/W^\pm$, which will be modified in the presence of new physics that modifies $Z\bar ff/Wff'$ couplings and/or introduces contact 4-fermion interactions. Specific ultraviolet examples include the $Z'$ model or a neutral scalar $\phi$ in the neutral current case, and an intermediate $W_R$ exchange from the left-right symmetric model for example or a charged scalar $\phi^\pm$ in the charged current case. Here in this work, we are aiming at a model independent analysis of \neff in the SMEFT, thus we will avoid any specific model discussion in the following even though this extension is straightforward.

\subsection{\boldmath From the SMEFT to \neff}\label{subsec:opes}
As mentioned in last subsection, if new physics modifies $Z\bar ff/Wff'$ couplings and/or introduces effective 4-fermion interactions, the prediction of \neff will be modified in general. Examples include the $Z'$ model and the $W_R$ exchange from the left-right symmetric model discussed earlier. While these models may have very rich physics at colliders or low-energy experiments like neutrino oscillations, we will adopt the SMEFT framework in this work to avoid any bias in model selection. We shall point out that the results presented in this work can be directly translated onto specific models, provided the new physics is heavy to ensure the validity of the SMEFT, see \cite{deBlas:2022ofj} for example in this regard. In what follows, we will first review the SMEFT operators relevant for \neffe, and then discuss their impact on the latter.

\subsubsection{\boldmath The SMEFT framework}
In the SMEFT, we parameterize the electroweak vertices and the contact 4-fermion interactions in the Higgs basis\,\cite{LHCHiggsCrossSectionWorkingGroup:2016ypw}, and summarize our notations in eq.\,\eqref{eq:vertex} and Table\,\ref{tab:4fope}, respectively. {There $g_L$ and $g_Y$ represent the SU(2)$_L$ and U(1)$_Y$ gauge couplings, $f_I\,(f=e,\nu)$ and $e^c_I$ with $I=1,2,3$ are the left- and right-handed leptons following the 2-component spinor formalism in\,\cite{Dreiner:2008tw}. $T_3^f$ is the isospin projector for fermion $f$ whose electric charge is given by $Q_f$, and $s_W=g_Y/\sqrt{g_L^2 + g_Y^2}$ is the sine of the weak mixing angle.} Generically speaking, $Zqq$ and $Wqq'$ vertices are also modified by SMEFT operators and can be parameterized in a similar way as those in eq.\,\eqref{eq:vertex}. These quark vertices are however irrelevant for the calculation of \neff due to the approximation we adopt here and thus not included explicitly. Similar arguments apply to semi-leptonic 4-fermion operators, and we therefore only show the pure leptonic operators in Table\,\ref{tab:4fope}.
\begin{widetext}
\eqal{
{\cal L}_{Vff}  = \, & {g_L \over \sqrt 2}\left [ W^{\mu+}  \bar \nu_I \bar \sigma_\mu (\delta_{IJ} +  [\delta g^{W \ell}_L]_{IJ}  ) e_J + h.c. \right ] \nonumber\\
& + \sqrt{g_L^2 + g_Y^2} Z^\mu    
\sum_{f=e,\nu}\bar f_I \bar \sigma_\mu \left (  (T_3^f -s_W^2 Q_f) \delta_{IJ} + \left[\delta g^{Zf}_L\right]_{IJ} \right ) f_J \nonumber\\
& + \sqrt{g_L^2 + g_Y^2} Z^\mu e_I^c \sigma_\mu \left ( - s^2_W Q_e \delta_{IJ} + \left[\delta g^{Z e}_R \right]_{IJ}\right ) \bar e_J^c,\label{eq:vertex}
}
\end{widetext}

\begin{table}[!htb]\caption{4-fermion operators contributing to \neffe, {where $\ell$ represents the SU(2)$_L$ leptons, either neutral or charged, and $e^c$ the right-handed charged lepton singlet.}}\label{tab:4fope}
\centering{
\begin{tabular}{l|c}
\hline\hline
\multicolumn{2}{c}{4-fermion operators}\\
\hline
\hspace{.4cm}One flavor ($p,r = 1,2,3$) &  \hspace{.4cm}Two flavors ($p < r =1,2,3$)\\ 
\hline 
$ [\mathcal{O}_{ll}]_{pppp} = {1\over 2} (\bar \ell_p\bar \sigma_\mu \ell_p)  (\bar \ell_p \bar \sigma^\mu \ell_p)$  &  $ [\mathcal{O}_{ll}]_{pprr}  =  (\bar \ell_p\bar \sigma_\mu \ell_p)  (\bar \ell_r \bar \sigma^\mu \ell_r) $ \\ 
$ [\mathcal{O}_{l e}]_{pppp} =  (\bar \ell_p\bar \sigma_\mu \ell_p)  (e_p^c  \sigma^\mu \bar e_p^c) $
&  $[\mathcal{O}_{ll}]_{prrp} = (\bar \ell_p \bar \sigma_\mu \ell_r)  (\bar \ell_r \bar \sigma^\mu \ell_p)  $\\   
& $ [\mathcal{O}_{l e}]_{pprr}  =  (\bar \ell_p\bar \sigma_\mu \ell_p)  (e_r^c  \sigma^\mu \bar e_r^c)$\\
& $[\mathcal{O}_{l e}]_{rrpp}  =  (\bar \ell_r \bar \sigma_\mu \ell_r)  (e_p^c  \sigma^\mu \bar e_p^c)$ \\
& $[\mathcal{O}_{l e}]_{prrp}  =  (\bar \ell_p \bar \sigma_\mu \ell_r)  (e_r^c  \sigma^\mu \bar e_p^c)$ \\ 
\hline\hline
\end{tabular}
}
\end{table}

Several comments are in order:
\begin{itemize}
\item One extra 4-lepton operator $[\mathcal{O}_{ee}]_{1111}$ is present in the SMEFT. This operator will contribute to electron scattering/annihilation during neutrino decoupling, which in general also depends on $[\mathcal{O}_{ll,l e}]_{1111}$. The Bhabha process at LEP\,\cite{ALEPH:2013dgf} with different energy runs and the M\o ller process at SLAC-E158\,\cite{SLACE158:2005uay}, as well as the next generation parity-violating electron scattering experiment MOLLER at the Jefferson Laboratory\,\cite{MOLLER:2014iki} with polarized electron beams, can be used to stringently constrain them. For this study, since $[\mathcal{O}_{ee}]_{1111}$ neither changes the number nor the energy density of electrons, it will not modify \neff and can therefore be neglected.
\item The two 4-fermion operators in the left column contribute to $\nu_e$-$e$ and $\nu_e$-$\nu_e$ scattering/annihilation.\footnote{Scattering/Annihilation involving anti-particles is to be understood with our notations here and in the following.} The first process had been measured in the past by several experiments including the Savannah River Plant\,\cite{Reines:1976pv}, LAMPF\,\cite{Allen:1992qe}, CHARM\,\cite{CHARM-II:1994aeb,CHARM-II:1995xfh}, LSND\,\cite{LSND:2001akn}, and TEXONO\,\cite{Deniz:2010mp}, as nicely reviewed in \cite{Formaggio:2012cpf}. The latter process is more challenging from homemade targets, but the cosmological probe \neff can be used to provide this missing information.
\item The remaining 4-fermion operators in the right column of Table\,\ref{tab:4fope} would contribute to $\nu_{\mu,\tau}$ scattering/annihilation. Depending on the strength of the corresponding Wilson coefficients, they can either delay or advance the decoupling of neutrinos from the plasma, thus modifying the prediction of \neff at late time. In turn, a precision measurement of \neff from Planck\,\cite{Aghanim:2018eyx} can be used to constrain these operators, which will be further improved by next-generation experiments introduced in the previous section, and will be the main focus of this work.
\item Similar to the 4-fermion operators, electroweak vertices in eq.\,\eqref{eq:vertex} can also change the prediction of \neff through modifying the interacting strengths between the SU(2) gauge bosons and the SM fermions. These vertices are stringently constrained at the $10^{-3}\sim10^{-4}$ level currently\,\cite{deBlas:2022ofj}, such that one can safely ignore their effects on \neff during the global fit. For this study, we nevertheless include them in our analysis, and confirm that their impact on \neff is marginal, as will become clear in section\,\ref{sec:results}.
\item Specific flavor assumption can be made to reduce the number of parameters and thus simplify the analysis. We will not make any flavor assumption in this work in order to be generic. Generalization to specific preferred flavor scenarios will be straightforward at the end of the analysis, and global fit results in the flavor universal case will be presented for illustration.
\item It is common in literature to assume the indistinguishability between $\nu_\mu$ and $\nu_\tau$ for the calculation of \neff during the evolution of the Universe, i.e., completely ignoring the flavor difference between $\nu_\mu$ and $\nu_\tau$. This is, however, in contradiction with the spirit of this work, viz, a flavor general global fit of electroweak vertices and 4-fermion operators. For this reason, we will track independently the evolution of $T_{\gamma,\nu_e,\nu_\mu,\nu_\tau}$ for a precision calculation of \neff in the SMEFT, which is one of the novel aspects of this work.\footnote{{This does \textit{not} mean we assume $T_{\nu_\mu}\ne T_{\nu_\tau}$ in practice but only that we consider the flavor difference between them and solve their temperature evolution individually for $\nu_\mu$ and $\nu_\tau$ with identical initial conditions. In the end of the numerical solution, we find $T_{\nu_\mu} = T_{\nu_\tau}$ is always maintained and consistent with constraints from neutrino oscillations.}} As a result, as we shall see shortly, we are able to probe more SMEFT operators in the $\tau$ sector utilizing \neffe.
\end{itemize}

With this, we will then calculate SMEFT corrections to the collision term integrals in the next subsection. {Before presenting the details of SMEFT corrections to \neffe, here we also briefly comment on the difference between this work and that in \cite{Du:2021idh}. This difference basically lies in the following three aspects: (1) Ref.\,\cite{Du:2021idh} considered the neutral current neutrino non-standard interactions in the low-energy EFT (LEFT) framework below the weak scale, while in this work we will utilize the SMEFT framework above the weak scale with dramatically different operators; (2) Constraints on the LEFT operators in Ref.\,\cite{Du:2021idh} were obtained by considering one operator at a time, while this work will take the full correlations among different SMEFT operators into account for the reason detailed in the Introduction; (3) To obtain constraints on the SMEFT operators from a global analysis, this work will combine different data sets from various high- and low-energy experiments together with \neff from Planck, CMB-S4/HD. In contrast, those constraints on the LEFT operators in Ref.\,\cite{Du:2021idh} were obtained by only using \neff from Planck and CMB-S4.
}

\subsubsection{\boldmath Corrections to the collision terms from the SMEFT}
The invariant amplitude for each process in Table\,\ref{tab:procs} receives SMEFT corrections from eq.\,\eqref{eq:vertex} and Table\,\ref{tab:4fope}, which can be generically written as
\eqal{\langle\mathcal{M}^{2}\rangle &\, = \langle \mathcal{M}_{\rm SM}^{2}\rangle + 2\,{\rm Re} \langle\mathcal{M}_{\rm SM}\cdot\mathcal{M}_{\rm SMEFT}^\dagger\rangle + \langle \mathcal{M}_{\rm SMEFT}^{2}\rangle\nb\\
&\, \approx \langle \mathcal{M}_{\rm SM}^{2}\rangle + 2\,{\rm Re} \langle\mathcal{M}_{\rm SM}\cdot\mathcal{M}_{\rm SMEFT}^\dagger\rangle,\label{amp2Inter}}
where $\mathcal{M}_{\rm SM}$ and $\mathcal{M}_{\rm SMEFT}$ are the SM and the SMEFT amplitudes, respectively. As mentioned earlier, in this work, we will only keep the leading order corrections from the SMEFT operators, thus the last term in the first line of eq.\,\eqref{amp2Inter} will be neglected in the following.

For reference, we summarize the leading order SMEFT corrections to the invariant amplitudes of each process in Table\,\ref{tab:procs} in this subsection. In each case, the amplitudes are given in the same order as that in each column of Table\,\ref{tab:procs}, with the symmetry factor implicitly included in each expression, $G_F$ the Fermi constant, $m_e$ the electron mass, and $p_{ij}\equiv p_i\cdot p_j$ being the scalar product of any two four-momenta.

\begin{widetext}
\begin{itemize}
\item $\nu_e$ case ($j=\mu,\tau$, or equivalently, $j=2,3$):
\eqal{
\bra\mathcal{M}^2\ket_1 &\, = 128 G_F^2 \cdot p_{14}^2 \cdot \left(1 - 2 [c_{ll}]_{1111} + 8 \delta g_{W}^{e\nu} + 8 \delta g_{Z,L}^{ee}\right),\label{eq:ve1}
}
\eqal{
\bra\mathcal{M}^2\ket_2 &\, = 64 G_F^2 \cdot p_{12}^2 \cdot \left(1 - 2 [c_{ll}]_{1111} + 8 \delta g_{W}^{e\nu} + 8 \delta g_{Z,L}^{ee}\right),\\
\bra\mathcal{M}^2\ket_3^j &\, = 32 G_F^2 \cdot p_{14}^2 \cdot \left(1 - 2 [c_{ll}]_{11jj} - 2 [c_{ll}]_{1jj1} + 4 \delta g_{W}^{e\nu} + 4 \delta g_{W}^{j\nu_j} + 4 \delta g_{Z,L}^{ee} + 4 \delta g_{Z,L}^{jj} \right),\label{eq:vecase3}\\
\bra\mathcal{M}^2\ket_4^j &\, = \bra\mathcal{M}^2\ket_3^j, \\
\bra\mathcal{M}^2\ket_5^j &\, = \frac{p_{12}^2}{p_{14}^2} \cdot \bra\mathcal{M}^2\ket_3^j,\\
\bra\mathcal{M}^2\ket_6 &\, = 32 G_F^2 \left[2 m_e^2 s_W^2 (1+2s_W^2)p_{12} + (1+2s_W^2)^2 p_{13}^2 + 4 s_W^4 p_{14}^2 - 2 [c_{ll}]_{1111} (p_{13}^2 + (m_e^2 p_{12} + 2 p_{13}^2) s_W^2)\right.\nb\\
&\left.\qquad\qquad\quad - [c_{le}]_{1111} (4 p_{14}^2 s_W^2 + m_e^2 p_{12} (1 + 2 s_W^2)) + 4 (p_{13}^2 + 2 (m_e^2 p_{12} + 2 p_{13}^2) s_W^2\right.\nb\\
&\left.\qquad\qquad\quad + 4 (m_e^2 p_{12} + p_{13}^2 + p_{14}^2) s_W^4) \delta g_{W}^{e\nu} + 4 s_W^2 (m_e^2 p_{12} + 2 p_{13}^2 + 4 (m_e^2 p_{12} + p_{13}^2 + p_{14}^2) s_W^2) \delta g_{Z,L}^{ee}\right.\nb\\
&\left.\qquad\qquad\quad + (8 s_W^2 p_{14}^2 + 2 m_e^2 (1 + 2 s_W^2)p_{12}) \delta g_{Z,R}^{ee} \right],\\
\bra\mathcal{M}^2\ket_7 &\, = 32 G_F^2 \left[(1+2s_W^2)^2 p_{12}^2 - 2 m_e^2 s_W^2(1+2s_W^2) p_{13} + 4s_W^4 p_{14}^2 - 2 [c_{ll}]_{1111} (p_{12}^2 + (2 p_{12}^2 - m_e^2 p_{13}) s_W^2)\right.\nb\\
&\left.\qquad\qquad\quad - [c_{le}]_{1111} (4 p_{14}^2 s_W^2 - m_e^2 p_{13} (1 + 2 s_W^2)) + 4 s_W^2 (2 p_{12}^2 - m_e^2 p_{13} + 4 (p_{12}^2 -m_e^2 p_{13} + p_{14}^2) s_W^2) \delta g_{Z,L}^{ee}\right.\nb\\
&\left.\qquad\qquad\quad + 4 (p_{12}^2 + 2 (2 p_{12}^2 - m_e^2 p_{13}) s_W^2 + 4 (p_{12}^2 - m_e^2 p_{13} + p_{14}^2) s_W^4) \delta g_{W}^{e\nu}\right.\nb\\
&\left.\qquad\qquad\quad + (8 s_W^2 p_{14}^2 - 2 m_e^2 (1 + 2 s_W^2)p_{13}) \delta g_{Z,R}^{ee} \right],\\
\bra\mathcal{M}^2\ket_8 &\, = 32 G_F^2 \left[4s_W^4 p_{12}^2 - 2 m_e^2 s_W^2(1+2s_W^2) p_{13} + (1+2s_W^2)^2 p_{14}^2 - 2 [c_{ll}]_{1111} (p_{14}^2 + (2 p_{14}^2 - m_e^2 p_{13}) s_W^2)\right.\nb\\
&\left.\qquad\qquad\quad - [c_{le}]_{1111} (4 p_{12}^2 s_W^2 - m_e^2 p_{13} (1 + 2 s_W^2)) + 4 s_W^2 (2 p_{14}^2 - m_e^2 p_{13}  + 4 (p_{12}^2 - m_e^2 p_{13} + p_{14}^2) s_W^2) \delta g_{Z,L}^{ee}\right.\nb\\
&\left.\qquad\qquad\quad + 4 (p_{14}^2 + 2 (2 p_{14}^2 - m_e^2 p_{13}) s_W^2 + 4 (p_{12}^2 - m_e^2 p_{13} + p_{14}^2) s_W^4) \delta g_{W}^{e\nu}\right.\nb\\
&\left.\qquad\qquad\quad + (8 s_W^2 p_{12}^2 - 2 m_e^2 (1 + 2 s_W^2)p_{13}) \delta g_{Z,R}^{ee} \right],
}
\item $\nu_j$ ($j=\mu,\tau$, or equivalently, $j=2,3$, respectively) case:
\hspace{-3cm}\eqal{%
\bra\mathcal{M}^2\ket_1^j &\, = 128 G_F^2 \cdot p_{14}^2 \cdot \left(1 - 2 [c_{ll}]_{jjjj} + 8 \delta g_{W}^{j\nu_j} + 8 \delta g_{Z,L}^{jj}\right),\label{eq:vuj1}\\
\bra\mathcal{M}^2\ket_2 &\, = 64 G_F^2 \cdot p_{12}^2 \cdot \left(1 - 2 [c_{ll}]_{jjjj} + 8 \delta g_{W}^{j\nu_j} + 8 \delta g_{Z,L}^{jj}\right),\label{eq:vuj2}\\
\bra\mathcal{M}^2\ket_3 &\, = 32 G_F^2 \cdot p_{14}^2 \cdot \left(1 - 2 [c_{ll}]_{11jj} - 2 [c_{ll}]_{1jj1} + 4 \delta g_{W}^{e\nu} + 4 \delta g_{W}^{j\nu_j} + 4 \delta g_{Z,L}^{ee} + 4 \delta g_{Z,L}^{jj} \right),\quad \text{for }(\nu_j\bar\nu_j\to\nu_e\bar\nu_e)\\
\bra\mathcal{M}^2\ket_3 &\, = 32 G_F^2 \cdot p_{14}^2 \cdot \left(1 - 2 [c_{ll}]_{2233} - 2 [c_{ll}]_{2332} + 4 \delta g_{W}^{\mu\nu} + 4 \delta g_{W}^{\tau\nu} + 4 \delta g_{Z,L}^{\mu\mu} + 4 \delta g_{Z,L}^{\tau\tau} \right),\quad \text{for }(\nu_j\bar\nu_j\to\nu'_j\bar\nu'_{j})\\
\bra\mathcal{M}^2\ket_4^j &\, = \bra\mathcal{M}^2\ket_3^j,\\
\bra\mathcal{M}^2\ket_5^j &\, = \frac{p_{12}^2}{p_{14}^2} \cdot \bra\mathcal{M}^2\ket_3^j,\\
\bra\mathcal{M}^2\ket_6^j &\, = 32 G_F^2 \left[2 m_e^2 s_W^2 (2s_W^2 - 1)p_{12} + (1- 2s_W^2)^2 p_{13}^2 + 4 s_W^4 p_{14}^2 + [c_{ll}]_{11jj} (2p_{13}^2 - 2(m_e^2 p_{12} + 2 p_{13}^2) s_W^2)\right.\nb\\
&\left.\qquad\qquad\quad - [c_{le}]_{jj11} (4 p_{14}^2 s_W^2 - m_e^2 p_{12} (1 - 2 s_W^2)) - 4 (p_{13}^2 - (m_e^2 p_{12} + 2 p_{13}^2) s_W^2) \delta g_{Z,L}^{ee}\right.\nb\\
&\left.\qquad\qquad\quad + 4 (p_{13}^2 - 2 (m_e^2 p_{12} + 2 p_{13}^2) s_W^2 + 4 (m_e^2 p_{12} + p_{13}^2 + p_{14}^2) s_W^4) \delta g_{W}^{j\nu_j}\right.\nb\\
&\left.\qquad\qquad\quad + 4 (p_{13}^2 - 2 (m_e^2 p_{12} + 2 p_{13}^2) s_W^2 + 4 (m_e^2 p_{12} + p_{13}^2 + p_{14}^2) s_W^4) \delta g_{Z,L}^{jj} \right.\nb\\
&\left.\qquad\qquad\quad + (8 s_W^2 p_{14}^2 - 2 m_e^2 (1 - 2 s_W^2)p_{12}) \delta g_{Z,R}^{ee} \right],\\
\bra\mathcal{M}^2\ket_7^j &\, = 32 G_F^2 \left[(1- 2s_W^2)^2 p_{12}^2 + 2 m_e^2 s_W^2 (1 - 2s_W^2)p_{13} + 4 s_W^4 p_{14}^2 + 2 [c_{ll}]_{11jj} (p_{12}^2 + m_e^2 s_W^2 p_{13} - 2 p_{12}^2 s_W^2)\right.\nb\\
&\left.\qquad\qquad\quad - [c_{le}]_{jj11} (4 p_{14}^2 s_W^2 + m_e^2 p_{13} (1 - 2 s_W^2)) - 4 (p_{12}^2 + m_e^2 p_{13} s_W^2 - 2 p_{12}^2 s_W^2) \delta g_{Z,L}^{ee}\right.\nb\\
&\left.\qquad\qquad\quad + 4 (p_{12}^2 + 2 (m_e^2 p_{13} - 2 p_{12}^2) s_W^2 + 4 (p_{12}^2 - m_e^2 p_{13} + p_{14}^2) s_W^4) \delta g_{W}^{j\nu_j}\right.\nb\\
&\left.\qquad\qquad\quad + 4 (p_{12}^2 + 2 (m_e^2 p_{13} - 2 p_{12}^2) s_W^2 + 4 (p_{12}^2 - m_e^2 p_{13} + p_{14}^2) s_W^4) \delta g_{Z,L}^{jj} \right.\nb\\
&\left.\qquad\qquad\quad + (8 s_W^2 p_{14}^2 + 2 m_e^2 (1 - 2 s_W^2)p_{13}) \delta g_{Z,R}^{ee} \right],
}
\eqal{
\bra\mathcal{M}^2\ket_8^j &\, = 32 G_F^2 \left[ 4 s_W^4 p_{12}^2 + 2 m_e^2 s_W^2 (1 - 2s_W^2)p_{13} + (1- 2s_W^2)^2 p_{14}^2 + 2 [c_{ll}]_{11jj} (p_{14}^2 + m_e^2 s_W^2 p_{13} - 2 p_{14}^2 s_W^2)\right.\nb\\
&\left.\qquad\qquad\quad - [c_{le}]_{jj11} (4 p_{12}^2 s_W^2 + m_e^2 p_{13} (1 - 2 s_W^2)) - 4 (p_{14}^2 + m_e^2 p_{13} s_W^2 - 2 p_{14}^2 s_W^2) \delta g_{Z,L}^{ee}\right.\nb\\
&\left.\qquad\qquad\quad + 4 (p_{14}^2 + 2 (m_e^2 p_{13} - 2 p_{14}^2) s_W^2 + 4 (p_{12}^2 - m_e^2 p_{13} + p_{14}^2) s_W^4) \delta g_{W}^{j\nu_j}\right.\nb\\
&\left.\qquad\qquad\quad + 4 (p_{14}^2 + 2 (m_e^2 p_{13} - 2 p_{14}^2) s_W^2 + 4 (p_{12}^2 - m_e^2 p_{13} + p_{14}^2) s_W^4) \delta g_{Z,L}^{jj} \right.\nb\\
&\left.\qquad\qquad\quad + (8 s_W^2 p_{12}^2 + 2 m_e^2 (1 - 2 s_W^2)p_{13}) \delta g_{Z,R}^{ee} \right],\label{eq:vjl}
}
\end{itemize}
\end{widetext}

In each case, we reproduce the corresponding SM result when setting all the Wilson coefficients to zero. Then, with the results above, the collision term integrals can be directly obtained from the dictionary derived in \cite{Du:2021idh}, which will serve as the input for us to numerically solve $T_\gamma$ and $T_{\nu_\alpha}$ before obtaining \neff through eq.\,\eqref{eq:nefffinal}. As an example, we consider SMEFT corrections to $ \nu_e + \bar\nu_e \to \nu_j + \bar\nu_j\,(j=\mu,\tau)$. The invariant amplitudes are given in eq.\,\eqref{eq:vecase3}. Then, from the dictionary in \cite{Du:2021idh}, their eq.(4.19) to be more specific, one immediately finds, for the number and the energy density respectively,
\begin{widetext}
\eqal{
\frac{\delta n_{\nu_e}}{\delta t} = &\, \frac{4G_F^2}{\pi^{5}}\left[-{e}^{\frac{2\mu_{\nu_e}}{T_{\nu_e}}} T_{\nu_e}^{8} + {e}^{\frac{2\mu_{\nu_j}}{T_{\nu_j}}} T_{\nu_j}^{8}\right] \cdot \left(1 - 2 [c_{ll}]_{11jj} - 2 [c_{ll}]_{1jj1} + 4 \delta g_{W}^{e\nu} + 4 \delta g_{W}^{j\nu_j} + 4 \delta g_{Z,L}^{ee} + 4 \delta g_{Z,L}^{jj} \right),\\
\frac{\delta \rho_{\nu_e}}{\delta t} = &\, \frac{4G_F^2}{\pi^{5}}\left[-4{e}^{\frac{2\mu_{\nu_e}}{T_{\nu_e}}} T_{\nu_e}^{9} + 4{e}^{\frac{2\mu_{\nu_j}}{T_{\nu_j}}} T_{\nu_j}^{9}\right] \cdot \left(1 - 2 [c_{ll}]_{11jj} - 2 [c_{ll}]_{1jj1} + 4 \delta g_{W}^{e\nu} + 4 \delta g_{W}^{j\nu_j} + 4 \delta g_{Z,L}^{ee} + 4 \delta g_{Z,L}^{jj} \right).
}
\end{widetext}

The same procedure is applied to all the processes in Table\,\ref{tab:procs} using the invariant amplitudes in eqs.\,(\ref{eq:ve1}-\ref{eq:vjl}), and keeping the leading order SMEFT corrections to $T_\gamma$ and $T_{\nu_\alpha}$ only. Then we numerically solve eqs.\,(\ref{TgammaFinal}-\ref{chemevolv}) to get \neff from eq.\,\eqref{eq:nefffinal}, which we discuss in the following subsections.
\subsubsection{\boldmath Numerical strategy to \neff calculation within the SMEFT}\label{sec:eft2neffstra}
In the presence of SMEFT operators, the differential equations in eqs.\,(\ref{TgammaFinal}-\ref{chemevolv}) can be generically written as
\begin{widetext}
\eqal{
\frac{d T_{\gamma}}{d t} = &\, f_0(T_\gamma,T_{\nu_\alpha}, \mu_{\nu_\alpha}, g_{SM}) + \sum_i c_i\, f_i (T_\gamma,T_{\nu_\alpha}, \mu_{\nu_\alpha}, g_{SM}),\label{eq:dTadt} \\
\frac{d T_{\nu_{\alpha}}}{d t} =&\, g_0(T_\gamma,T_{\nu_\alpha}, \mu_{\nu_\alpha}, g_{SM}) + \sum_i c_i\, g_i (T_\gamma,T_{\nu_\alpha}, \mu_{\nu_\alpha}, g_{SM}),\label{eq:dTvdt} \\
\frac{d \mu_{\nu_{\alpha}}}{d t} =&\, h_0(T_\gamma,T_{\nu_\alpha}, \mu_{\nu_\alpha}, g_{SM}) + \sum_i c_i\, h_i (T_\gamma,T_{\nu_\alpha}, \mu_{\nu_\alpha}, g_{SM}), \label{eq:dudt}
}
\end{widetext}
where $f_{0}$, $g_{0}$, and $h_0$ include contributions from {\textit{both}} the scattering/annihilation processes in the SM {\textit{and}} the dilution effect due to the expanding Universe, and are functions of the photon and neutrino temperatures, and the neutrino chemical potentials. $f_{i}$, $g_{i}$, and $h_i$ account for the extra scattering/annihilation corrections from the SMEFT, where $c_i$'s are the Wilson coefficients of the vertex and 4-fermion types introduced earlier. $g_{SM}$ represents the SM input parameters, and the summation is over the SMEFT operators. Solutions to the above differential equations can be generically written as
\eqal{
T_{\gamma} = T_{\gamma} & (g_{SM},c_i), \quad T_{\nu_{\alpha}} = T_{\nu_{\alpha}}(g_{SM},c_i),\nb\\
&\, \mu_{\nu_{\alpha}} = \mu_{\nu_{\alpha}}(g_{SM},c_i),\label{eq:tempsol}
}
which are expressed as parametric functions of the Wilson coefficients after solving eqs.\,(\ref{eq:dTadt}-\ref{eq:dudt}) numerically. These parametric functions are then used to compute \neff as follows:
\eqal{
N_{\rm eff} &\, = N_{\rm eff,\,SM} + \sum_i c_i\, k_i + \sum_{i ,j} c_i c_j \, l_{i,j} + \cdots \nb\\
&\, \approx N_{\rm eff,\,SM} + \sum_i c_i\, k_i,\label{eq:nefflinear}
}
where the first term $N_{\rm eff,\,SM}$ is the SM prediction of \neffe, and the following terms corrections from the SMEFT at the linear, quadratic and higher orders. Due to the smallness of $c_i$'s, we only keep terms linear in them in the final step, and determine $N_{\rm eff,\,SM}$ and $k_i$'s from
\eqal{
N_{\rm eff,\,SM} = N_{\rm eff}(\{c\}=0), \quad k_i = \left.\frac{\partial N_{\rm eff}}{\partial c_i}\right\vert_{\{c\}=0},
}
where $\{c\}$ represents the collection of Wilson coefficients considered in this work.\footnote{We also calculate a few of the $l_{i,j}$'s and find them to be either smaller than or of the same order as the $k_i$'s, which partially justifies our linear order approximation.}

\subsubsection{\boldmath Corrections to \neff from the SMEFT}\label{sec:eft2neff}
Including the electroweak vertex and the 4-fermion operator corrections discussed above, we use the {\tt Mathematica} code {\tt EFT2Neff} implemented in-house and based on the dictionary in\,\cite{Du:2021idh} and the setup code {\tt nudec\_BSM} of \cite{Escudero:2020dfa} but with a new ordinary differential equation solver appropriate for the SMEFT, to numerically solve \neff in the SMEFT following the strategy detailed in the last subsection in the $0.009{\rm\,MeV}<T_\gamma<10{\rm\,MeV}$ temperature range. In addition, as commented above, we do not make any flavor assumption when calculating \neff within the SMEFT, and assume that all three flavor neutrinos have the same initial conditions with $T_\gamma^0 = T_{\nu_\alpha}^0 =10\rm\,MeV$, $ \mu_\gamma = 0 = \mu_e$ since photons and electrons/positions are still tightly coupled, and initial $\mu_{\nu_\alpha}^0 = -10^{-3}\rm\,MeV$ whose impact on \neff turns out to be negligible.\footnote{Assuming a slightly different initial condition will only change the corresponding coefficients in eq.\,\eqref{eq:neffsmeft} or the direction \neff is exploring in the full parameter space.} Then to the linear order in the SMEFT Wilson coefficients, we obtain
\eqal{
N_{\rm eff} = &\, 3.044 - 0.006 [c_{le}]_{1111} - 0.017 [c_{ll}]_{1111} \nb\\
&\, + 0.058 \delta g_{W}^{e\nu} - 0.006 \delta g_{Z,L}^{ee} + 0.038 \delta g_{Z,R}^{ee} \nb\\
&\, + 0.007 \left( [c_{ll}]_{1122} - [c_{le}]_{2211} \right)\nb\\
&\, + 0.007 \left( [c_{ll}]_{1133} - [c_{le}]_{3311} \right) \nb\\
&\, + 0.015\left( \delta g_{W}^{\mu\nu} + \delta g_{Z,L}^{\mu\mu}\right) \nb\\
&\, + 0.015\left( \delta g_{W}^{\tau\nu} + \delta g_{Z,L}^{\tau\tau} \right) , \label{eq:neffsmeft}
}
where we truncate the numbers in the above expression at the fourth digit due to the fact that, in practice, we find the linear order approximation in eq.\,\eqref{eq:nefflinear} introduces a numerical relative uncertainty at the $\mathcal{O}(10^{-5})$ level. Note that the uncertainty at this level is already beyond the precision goal of CMB-S4/HD even for $\mathcal{O}(1)$ $c$'s and/or $\delta g$'s, so corrections to \neff from any operator at/beyond this order can be safely neglected. 

Several points regarding eq.\,\eqref{eq:neffsmeft} merit addressing:
\begin{itemize}
\item It is clear from the first line that the SM prediction $N_{\rm eff}=3.044$\,\cite{Akita:2020szl,Froustey:2020mcq} is reproduced when all SMEFT corrections are absent.
\item $[c_{ll}]_{1221,1331,2233,2332}$ barely modify \neff due to the reason following eq.\,\eqref{eq:neffsmeft}. In particular, contributions to \neff from $[c_{ll}]_{2233,2332}$ vanish due to the fact that $\nu_\mu$ and $\nu_\tau$ share identical thermodynamics,\footnote{Numerically, we find $k_{[c_{ll}]_{2233,2332}}$ of $\mathcal{O}(10^{-11})$, which are effectively zeros.} which is also why the coefficients are the same for the second and third generation neutrinos as seen in the last four lines of eq.\,\eqref{eq:neffsmeft}. Furthermore, the coefficients of $\delta g_{W}^{\mu\nu}$ and $\delta g_{Z,L}^{\mu\mu}$, and similarly for $\tau$, are equal since both currents become purely left-handed for $\nu_{\mu,\tau}$.
\item Even though the amplitudes depend on $[c_{ll}]_{2222,3333}$ as can be directly seen from eqs.\,(\ref{eq:vuj1}-\ref{eq:vuj2}), \neff is independent of them since they do not change the energy and number densities of $\nu_{\mu,\tau}$. In contrast, the dependence of \neff on $[c_{ll}]_{1111}$ and $[c_{le}]_{1111}$ comes from the fact these two operators can change the energy and number densities of $\nu_e$ even though they do not affect those of $e^\pm$.\footnote{Recall that $\mu^\pm$ and $\tau^\pm$ have already all decayed away during neutrino decoupling, $[c_{ll}]_{2222,3333}$ will only contribute to 4-neutrino interactions without changing the number and energy densities of $\nu_{\mu,\tau}$. For $[c_{ll}]_{1111}$ and $[c_{le}]_{1111}$, though they will not change the number and energy densities of $e^\pm$, they will modify those of $\nu_e$ through, for example, $\nu_e+\bar\nu_e \leftrightarrow e^++e^-$.} We also comment that, experimentally, $[c_{ll}]_{2233}$ can be measured at a future muon collider\,\cite{Ankenbrandt:1999cta,Forslund:2022xjq,deBlas:2022aow}, and $[c_{ll}]_{2222,3333}$ can be measured through $Z\to4\mu/\tau$ or $h\to4\mu/\tau$\,\cite{CMS:2022kdx} even though the current statistics in the $\tau$ channel is extremely limited.\footnote{\neff is free of $\left[c_{l e,ee}\right]_{2222,3333}$ for the reason that muons and taus do not survive during neutrino decoupling.}
\item The relative sign and magnitude of each term in eq.\,\eqref{eq:neffsmeft} can be most straightforwardly understood from $\delta \rho_{\nu_\alpha}/\delta t$ and $\delta n_{\nu_\alpha}/\delta t$, which directly modify the time evolution of $T_{\gamma,\nu_{\alpha}}$ and $\mu_{\nu_{\alpha}}$ as seen in eqs.\,(\ref{TgammaFinal}-\ref{chemevolv}). For example, the total neutrino number density changing rate is given by
\begin{widetext}
\eqal{
\frac{\delta n_{\nu}^{\rm tot}}{\delta t} = &\, \frac{8G_F^2}{\pi^5} \left[\left(8 s_W^4+4 s_W^2+1\right) \left(T_\gamma^8-T_{\nu_e}^8\right) + 2 \left(8 s_W^4-4 s_W^2+1\right) \left(T_\gamma^8-T_{\nu_\mu}^8\right) \right. \nb\\
&\left. \, -  4 [c_{le}]_{1111} s_W^2 \left(T_\gamma^8 - T_{\nu_e}^8\right) - 2 [c_{ll}]_{1111} \left(2 s_W^2+1\right) \left(T_\gamma^8-T_{\nu_e}^8\right)\right. \nb\\
&\left. \, + 4 \delta g_{W}^{e\nu} \left(8 s_W^4+4 s_W^2+1\right)\left(T_\gamma^8-T_{\nu_e}^8\right) \right.\nb\\
&\left. \, + 8 \delta g_{Z,L}^{ee} \left( \left(4 s_W^4+3s_W^2 - 1 \right) T_\gamma^8 - \left(4 s_W^2+1\right) s_W^2 T_{\nu_e}^8 - \left(2 s_W^2-1\right) T_{\nu_\mu}^8 \right) \right. \nb\\
&\left. \, + 8 \delta g_{Z,R}^{ee} s_W^2 \left(3T_\gamma^8-2T_{\nu_\mu}^8-T_{\nu_e}^8\right) \right. \nb\\
&\left. \, - 4 \left( [c_{le}]_{2211} + [c_{le}]_{3311} \right) s_W^2 \left(T_\gamma^8 - T_{\nu_\mu}^8\right) \right. \nb\\
&\left. \, +  2 \left( [c_{ll}]_{1122} + [c_{ll}]_{1133} \right) \left(1 - 2s_W^2\right) \left(T_\gamma^8-T_{\nu_\mu}^8\right) \right. \nb\\
&\left. \, + 4 \left( \delta g_{W}^{\mu\nu} + \delta g_{Z,L}^{\mu\mu} + \delta g_{W}^{\tau\nu} + \delta g_{Z,L}^{\tau\tau} \right) \left(8 s_W^4-4 s_W^2+1\right) \left(T_\gamma^8-T_{\nu_\mu}^8\right) \right],\label{eq:neffdndt}
}
\end{widetext}
\begin{figure*}
\begin{center}
\begin{minipage}{0.9\linewidth}
\includegraphics[scale=0.41]{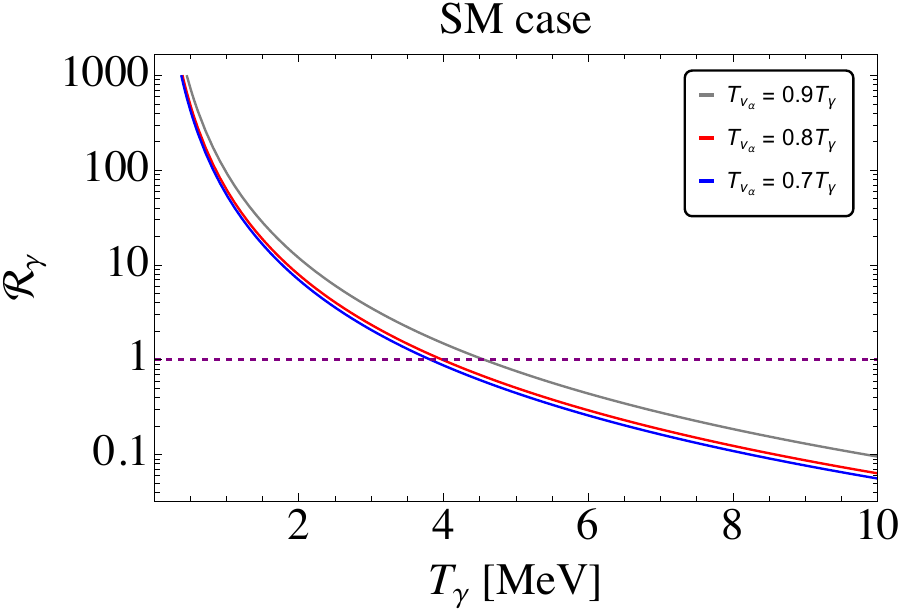} \quad\, \includegraphics[scale=0.4]{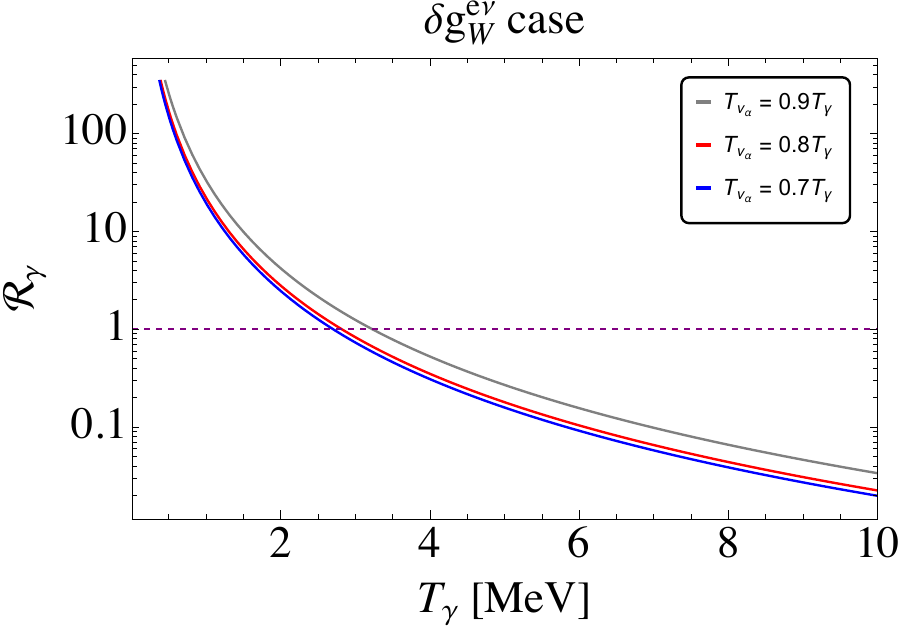} \\
\includegraphics[scale=0.41]{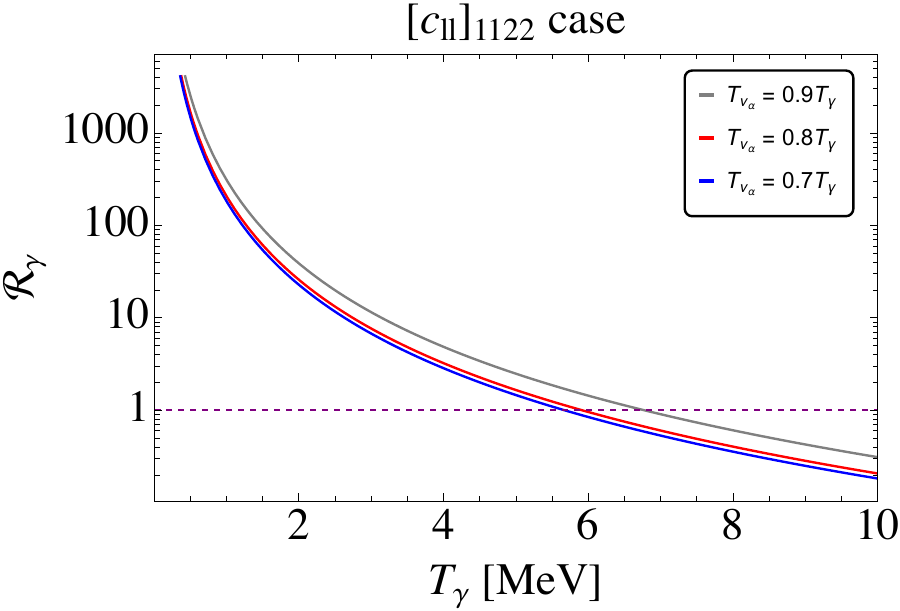} \quad \includegraphics[scale=0.411]{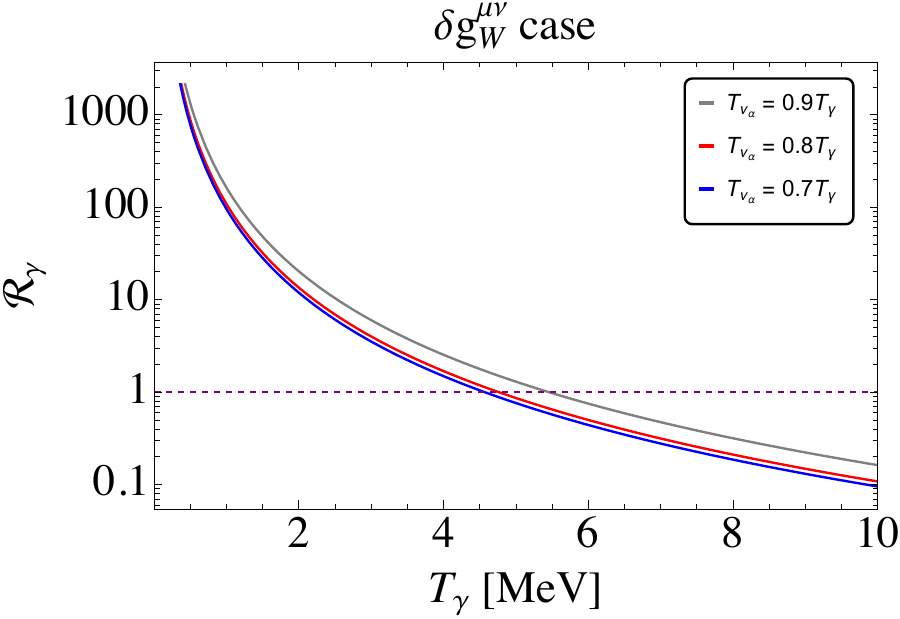}\caption{Plot for $\mathcal{R}_\gamma$ as a function of $T_\gamma$. The upper left plot is the ratio in the SM, and the others those for $\delta g_{W}^{e\nu}$, $[c_{ll}]_{1122}$ and $\delta g_{W}^{\mu\nu}$ as representatives of the SMEFT operators with larger, comparative, and smaller $\delta\rho^{\rm tot}_{\nu}/\delta t$ compared to the SM, \yong{obtained by fixing the respective Wilson coefficient at unity and the remaining ones at zero}. In each plot, different colors represent different neutrino temperatures as indicated by the legend, and the horizontal dashed line in purple is for $\mathcal{R}_\gamma=1$.}\label{fig:hubcolcompphoton}
\end{minipage}
\end{center}
\end{figure*}
\begin{figure*}
\begin{center}
\begin{minipage}{0.9\linewidth}
\includegraphics[scale=0.41]{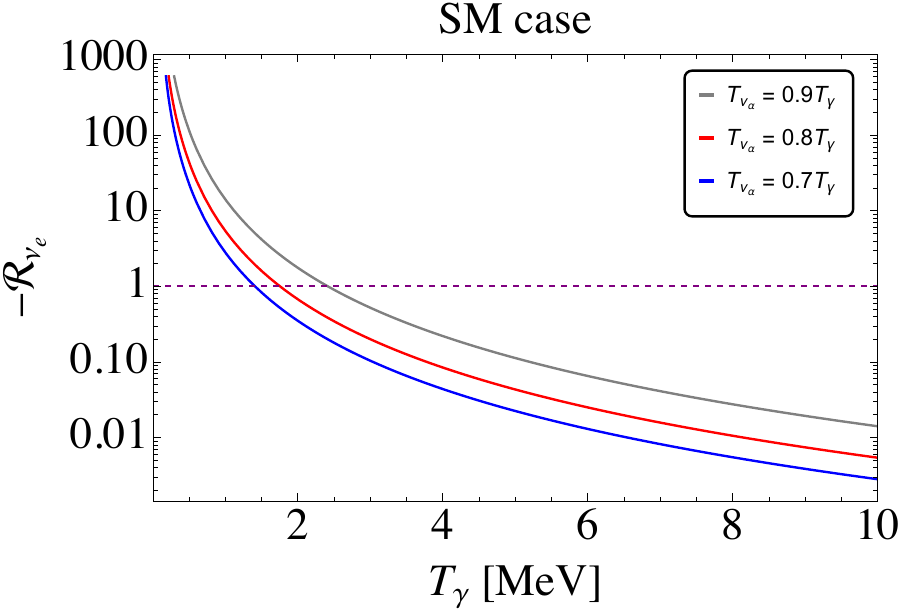} \,\,\, \includegraphics[scale=0.415]{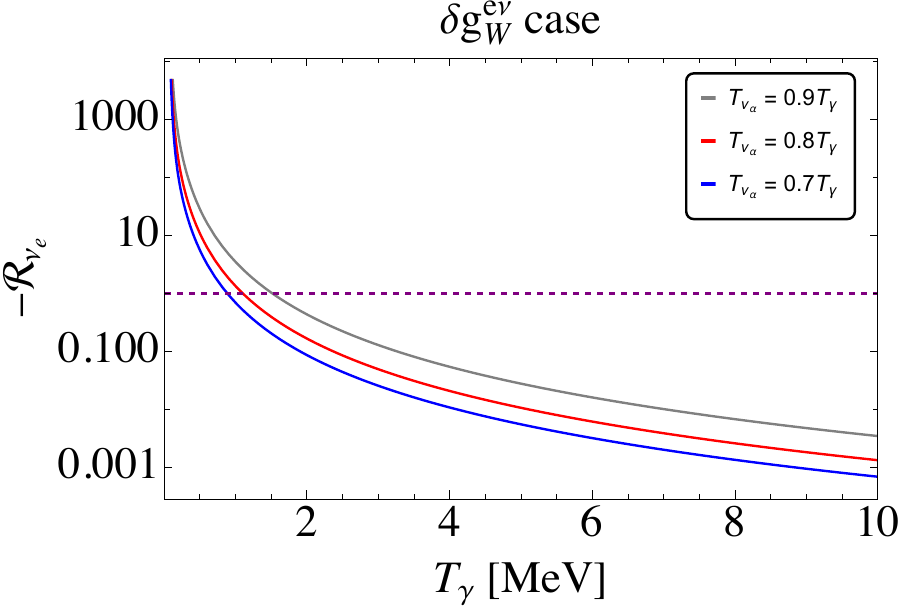} \\
\includegraphics[scale=0.41]{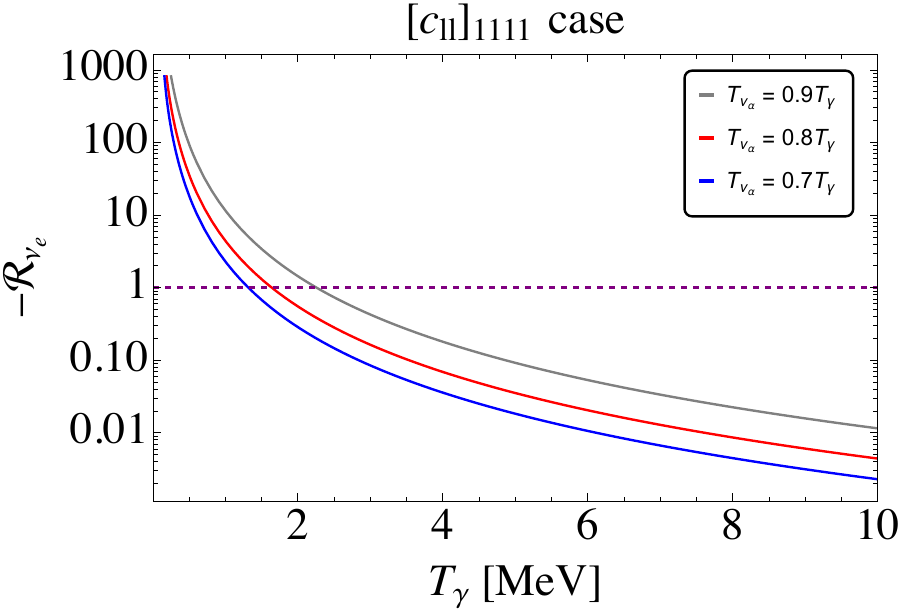} \,\,\,\,\, \includegraphics[scale=0.406]{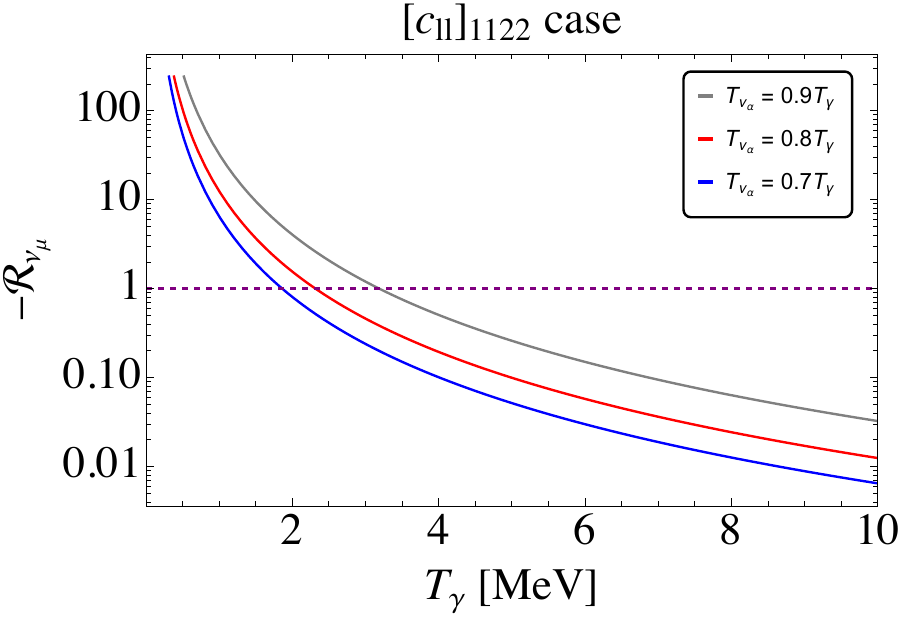} \caption{Same as figire\,\ref{fig:hubcolcompphoton} but for $\mathcal{R}_{\nu_\alpha}$. The negative sign in the vertical axis is simply to make the $y$-value positive for a logarithm plot. See the main text for details.}\label{fig:hubcolcompnue}
\end{minipage}
\end{center}
\end{figure*}
where for simplicity, we ignore small corrections from the finite electron mass and Fermi-Dirac statistics, and set $T_{\nu_\mu}=T_{\nu_\tau}$ and $\mu_{\nu_\alpha}=0$. The expression for $\delta \rho_{\nu}^{\rm tot}/\delta t$ is lengthy but very similar to ${\delta n_{\nu}^{\rm tot}}/{\delta t}$, which we show in Appendix\,\ref{app:drhodt} for reference. Recall that since $s_W^2\approx 1/4$, the second and third to last lines have approximately the same coefficients, which explains the coefficients in the second to last line of eq.\,\eqref{eq:neffsmeft}. Similarly, the accidental cancellation explains why the correction to \neff from $\delta g_{Z,L}^{ee}$ in eq.\,\eqref{eq:neffsmeft} is much smaller than the other operators.\footnote{Since $T_{\nu_e}\approx T_{\nu_\mu}=T_{\nu_\tau}$, the fourth line of eq.\,\eqref{eq:neffdndt} is approximately proportional to an overall factor $4 s_W^4+3s_W^2 - 1\approx 0$ due to accidental cancellation.} On the other hand, the signs of each term in eq.\,\eqref{eq:neffsmeft} can be understood from the fact that $T_\gamma>T_{\nu_\alpha}$ during the decoupling and the signs in eqs.\,(\ref{TgammaFinal}-\ref{chemevolv}). For example, the negative sign of the $[c_{le}]_{1111}$ term in eq.\,\eqref{eq:neffdndt} would effectively imply an increase in $T_\gamma$ and a decrease in $T_{\nu_e}$, thus a negative shift in \neffe, which is also consistent with eq.\,\eqref{eq:neffsmeft}.
\item The evolution of $T_\gamma$ is dominated by the expansion of the Universe, and that from neutrino collisions with the electromagnetic plasma is marginal during neutrino decoupling. This can be clearly seen from Fig.\,\ref{fig:hubcolcompphoton}, where the vertical axis $\mathcal{R}_\gamma$ represents the ratio of the expansion, plus the thermal effects, and the collision parts in eq.\,\eqref{TgammaFinal}:
\eqal{
\mathcal{R}_\gamma &\equiv \frac{4 H \rho_{\gamma}+3 H\left(\rho_{e}+p_{e}\right)+3 H T_{\gamma} \frac{d P_{\text {int }}}{d T_{\gamma}}}{\frac{\delta \rho_{\nu e}}{\delta t} + \frac{\delta \rho_{\nu \mu}}{\delta t} + \frac{\delta \rho_{\nu \tau}}{\delta t}}.
}
In each subplot of Fig.\,\ref{fig:hubcolcompphoton}, we set $\mu_{\nu_\alpha}=0$, $T_{\nu_e} = T_{\nu_\mu} = T_{\nu_\tau} = T_{\nu}$ for simplicity\yong{, and when applicable, we set all the Wilson coefficients to zero with the only surviving one at unity to obtain these plots.} Different colors in each plot represent variations of $T_\nu$ in terms of $T_\gamma$ as indicated by the legend. The very large $\mathcal{R}_\gamma$ during neutrino decoupling around $T_\gamma=2{\rm\,MeV}$ clearly justifies our claim about $T_\gamma$ evolution above. \yong{Though $T_\gamma$ is barely affected by the neutrinos, it is worth pointing out that depending on the specific operator present, the energy injection rate from the electromagnetic plasma into the neutrino sector could be different as can be seen by the representative examples shown in Fig.\,\ref{fig:hubcolcompphoton}. Specifically, this rate is enhanced, marginally affected, and reduced by $\delta g_W^{e\nu}$, $[c_{ll}]_{1122}$, and $\delta g_W^{\mu\nu}$, respectively, which can also be understood directly from eq.\,\eqref{eq:rhotot} from their relative magnitudes.}

\item Similarly, for the neutrinos, we define
\eqal{
\mathcal{R}_{\nu_\alpha} &\equiv -\frac{3 H\left[\left(\rho_{\nu_\alpha}+p_{\nu_\alpha}\right)\frac{\partial n_{\nu_\alpha}}{\partial \mu_{\nu_\alpha}} - n_{\nu_\alpha} \frac{\partial \rho_{\nu_\alpha}}{\partial \mu_{\nu_\alpha}} \right]}{\frac12\left[\frac{\partial n_{\nu_\alpha}}{\partial \mu_{\nu_\alpha}} \frac{\delta\rho_{\nu_\alpha}}{\delta t} - \frac{\partial \rho_{\nu_\alpha}}{\partial \mu_{\nu_\alpha}} \frac{\delta\rho_{\nu_\alpha}}{\delta t} \right] },
}
where $1/2$ in the denominator is to account for anti-neutrinos and the overall negative sign is to account for their relative signs. The corresponding plots for $\mathcal{R}_{\nu_\alpha}$ are shown in Fig.\,\ref{fig:hubcolcompnue}. From these plots, one can see that the evolution of $T_{\nu_\alpha}$ is different from that of $T_\gamma$ and critically depends on the relative contributions from both the expansion and the collisions, as expected. Note that for $\delta g_{W}^{e\nu}$ as shown in the upper right panel of Fig.\,\ref{fig:hubcolcompnue}, it will be the collisions rather than the expansion that mainly dominate the evolution of $T_{\nu_\alpha}$ relative to the SM scenario. As a result, the presence of this operator will tend to borrow more energy from the electromagnetic plasma to heat up the neutrinos, thus a larger shift of \neff as already seen in eq.\,\eqref{eq:neffsmeft}. \yong{Therefore, different from the evolution of $T_\gamma$ where the expansion and the thermal effects dominate, in obtaining the neutrino temperatures, neutrino collisions with the thermal plasma are significant in altering the energy rate from the plasma into the neutrino sector, especially in the presence of new physics as we have seen from $\delta g_{W}^{e\nu}$ for instance.}
\end{itemize}

With \neff in eq.\,\eqref{eq:neffsmeft}, we will next include it in the SMEFT global fit in the following section. Before presenting the results, we summarize the experimental inputs/projections we use for \neff in the next subsection.

\subsection{Experimental input parameters}\label{}
\begin{table}[!htb]\caption{\yong{Electroweak precision observables used in the global analysis, where ``Exp. values'' stand for experimental measurements, and numbers in the last column are from\,\cite{Baak:2014ora}}.}\label{tab:zwpoleobs}
\begin{center}
\begin{tabular}{|c|c|c|c|}
\hline
{Observables} & {Exp. values}   &   {Refs.}   &  {SM predictions}
  \\  \hline
$\Gamma_{Z}$ [GeV]  & $2.4952 \pm 0.0023$ & \cite{ALEPH:2005ab} & $ 2.4950$
 \\  \hline
$\sigma_{\rm had}$ [nb]  & $41.541\pm 0.037$ &\cite{ALEPH:2005ab} &  $41.484$
  \\  \hline
 $R_{e}$  & $20.804\pm 0.050$ & \cite{ALEPH:2005ab} &  $20.743$
   \\  \hline
 $R_{\mu}$  & $20.785 \pm 0.033$ & \cite{ALEPH:2005ab} &  $20.743$
   \\  \hline
 $R_{\tau}$  & $20.764\pm 0.045$ & \cite{ALEPH:2005ab} &  $20.743$
 \\  \hline
 $A_{\rm FB}^{0,e}$ & $0.0145\pm 0.0025$ &\cite{ALEPH:2005ab} &  $0.0163$
  \\  \hline
 $A_{\rm FB}^{0,\mu}$ & $0.0169\pm 0.0013$ &\cite{ALEPH:2005ab} &  $0.0163$
  \\  \hline
 $A_{\rm FB}^{0,\tau}$ & $0.0188\pm 0.0017$ &\cite{ALEPH:2005ab} &  $0.0163$
  \\ \hline
$R_b$ & $0.21629\pm0.00066$ & \cite{ALEPH:2005ab} & $0.21578$
   \\  \hline
$R_c$ & $0.1721\pm0.0030$  & \cite{ALEPH:2005ab}  & $0.17226$
\\  \hline
$A_{b}^{\rm FB}$ & $0.0992\pm 0.0016$ & \cite{ALEPH:2005ab}  & $0.1032$
 \\  \hline
 $A_{c}^{\rm FB}$ & $0.0707\pm 0.0035$  & \cite{ALEPH:2005ab} &  $0.0738$
  \\ \hline
 $A_e$ & $0.1516 \pm 0.0021$ &\cite{ALEPH:2005ab} &  $0.1472$
 \\ \hline
  $A_\mu$ & $0.142 \pm 0.015$ &\cite{ALEPH:2005ab} &  $0.1472$
 \\ \hline
 $A_\tau$ & $0.136 \pm 0.015$ &\cite{ALEPH:2005ab} &  $0.1472$
 \\ \hline
  $A_e$ & $0.1498 \pm 0.0049$ & \cite{ALEPH:2005ab} &  $0.1472$
 \\ \hline
 $A_\tau$ & $0.1439 \pm  0.0043$ & \cite{ALEPH:2005ab} &  $0.1472$
 \\ \hline
 $A_b$ & $0.923\pm 0.020$ & \cite{ALEPH:2005ab} & $0.935$
 \\  \hline
$A_c$ & $0.670 \pm 0.027$ & \cite{ALEPH:2005ab} & $0.668$
 \\ \hline
$A_s$ & $0.895 \pm 0.091$ & \cite{SLD:2000jop} & $0.935$
\\ \hline
$R_{uc}$ & $0.166 \pm 0.009$  & \cite{ParticleDataGroup:2012pjm}  & $0.1724$
\\ \hline
 $m_{W}$ [GeV]  & $80.385 \pm 0.015$ &\cite{Group:2012gb}    &  $80.364$
\\ \hline
$\Gamma_{W}$ [GeV]  & $ 2.085 \pm 0.042$  & \cite{ParticleDataGroup:2012pjm} &  $2.091$
\\ \hline
${\rm Br} (W \to e \nu)$ & $ 0.1071 \pm 0.0016$ &\cite{ALEPH:2013dgf} &  $0.1083$
\\ \hline
${\rm Br} (W \to \mu \nu)$ & $ 0.1063 \pm 0.0015$ &\cite{ALEPH:2013dgf} &  $0.1083$
\\ \hline
${\rm Br} (W \to \tau \nu)$ & $ 0.1138 \pm 0.0021$ &\cite{ALEPH:2013dgf} &  $0.1083$
 \\ \hline
 $R_{Wc}$ & $ 0.49 \pm 0.04$ & \cite{ParticleDataGroup:2012pjm}  &  $0.50$
 \\ \hline
 $R_{\sigma}$ & $0.998 \pm 0.041$  & \cite{CMS:2014mgj} & 1.000
  \\ \hline
\end{tabular}
\end{center}
\end{table}
\begin{table}\caption{\yong{Inputs from fermion pair production for the global analysis. The (differential) cross sections and asymmetries depend on the running energies and also the scattering angles. We refer the readers to the original tables of the cited references for the detailed numbers.}}\label{tab:fpairobs}
\begin{center}
\begin{adjustbox}{max width = \textwidth}
\begin{tabular}{|c|c|}
\hline
{Observables} & Refs.
 \\  \hline  \hline
$\sigma(\mu^+\mu^-)$ &  \cite{ALEPH:2013dgf}\\
\hline
$\sigma(\tau^+\tau^-)$ &  \cite{ALEPH:2013dgf}\\
\hline
$\sum\limits_{q\ne t}\sigma(q\bar{q})$ &  \cite{VENUS:1993pob,TOPAZ:2000evx,ALEPH:2013dgf}\\
\hline
$\sigma(b\bar{b})$ &  \cite{VENUS:1993pob,TOPAZ:2000evx,ALEPH:2006bhb}\\
\hline
$\sigma(c\bar{c})$ &  \cite{VENUS:1993pob,TOPAZ:2000evx,ALEPH:2006bhb}\\
\hline
${\sigma_{\rm FB}(b\bar{b})}/{(\sum\limits_{q\ne t}\sigma(q\bar{q}))}$ &  \cite{VENUS:1993pob,TOPAZ:2000evx,ALEPH:2006bhb}\\
\hline
${\sigma_{\rm FB}(c\bar{c})}/{(\sum\limits_{q\ne t}\sigma(q\bar{q}))}$ &  \cite{VENUS:1993pob,TOPAZ:2000evx,ALEPH:2006bhb}\\
\hline
$A_{\rm FB}(\mu^+\mu^-)$ &  \cite{ALEPH:2013dgf}\\
\hline
$A_{\rm FB}(\tau^+\tau^-)$ &  \cite{ALEPH:2013dgf}\\
\hline
${d\sigma}/{d\cos\theta}(\rm Bhabha)$ &  \cite{ALEPH:2013dgf}
\\ \hline
\end{tabular}
\end{adjustbox}
\end{center}
\end{table}
\begin{table}\caption{\yong{Low-energy observables included in the global analysis to lift possible flat directions in the SMEFT parameter space. We also include $A_{\rm FB}$ from the Drell-Yan process at the LHC in this table as indicated by the last row, for which we choose the 4 rapidity bins as in\,\cite{Breso-Pla:2021qoe}}.}\label{tab:leobs}
\centering
\begin{adjustbox}{max width = 0.5\textwidth}
\begin{tabular}{|c|c|c|c|c|}
\hline
{Processes} & {Observables} & {Experimental values}   &   {Refs.}   &  {SM predictions}
 \\  \hline  \hline
{$\stackrel{(-)}{\nu}_\mu-e^-$} & $g_{LV}^{\nu_\mu e}$ & $-0.035\pm0.017$ &\multirow{2}{*}{\cite{CHARM-II:1994dzw}}    &  $-0.0396$\cite{Erler:2013xha}\\
scattering & $g_{LA}^{\nu_\mu e}$ & $-0.503\pm0.017$  &   &  $-0.5064$\cite{Erler:2013xha}\\\hline
\multirow{2}{*}{$\tau$ decay} & $\frac{G_{\tau e}^2}{G_F^2}$ & $1.0029\pm0.0046$ & \multirow{2}{*}{\cite{ParticleDataGroup:2014cgo}}    &  \multirow{2}{*}{$1$}\\ 
 & $\frac{G_{\tau \mu}^2}{G_F^2}$ & $0.981\pm0.018$  &  & \\\hline
{} & $R_{\nu_\mu}$ & $0.3093\pm0.0031$ & \multirow{2}{*}{\cite{CHARM:1987pwr}} & 0.3156\cite{CHARM:1987pwr} \\
 & $R_{\bar{\nu}_\mu}$ & $0.390\pm0.014$ & & 0.370\cite{CHARM:1987pwr} \\
 \cline{2-5}
$\nu$ & $R_{\nu_\mu}$ & $0.3072\pm0.0033$ & \multirow{2}{*}{\cite{Blondel:1989ev}}  & 0.3091\cite{Blondel:1989ev}\\
scattering & $R_{\bar{\nu}_\mu}$ & $0.382\pm0.016$ & & 0.380\cite{Blondel:1989ev}\\
 \cline{2-5}
  & $\kappa$ & $0.5820\pm0.0041$ & \cite{CCFR:1997zzq} & 0.5830\cite{CCFR:1997zzq}\\
 \cline{2-5}
  & $R_{\nu_e\bar{\nu}_e}$ & $0.406^{+0.145}_{-0.135}$ & \cite{CHARM:1986vuz} & 0.33\cite{ParticleDataGroup:2016lqr}\\
\hline
 & $s^2_w$ & $0.2397\pm0.0013$ &\cite{SLACE158:2005uay}    &  $0.2381$\cite{Czarnecki:1995fw}\\
 \cline{2-5}
 & $Q_W^{\rm Cs}$ & $-72.62\pm0.43$ & \cite{ParticleDataGroup:2016lqr} & $-73.25$\cite{ParticleDataGroup:2016lqr} \\
 \cline{2-5}
 & $Q_W^{\rm p}$ & $0.064\pm0.012$ & \cite{Qweak:2013zxf} & $0.0708$\cite{ParticleDataGroup:2016lqr} \\
 \cline{2-5}
PV & $A_1 [10^{-6}]$ & $-91.1\pm4.3$ & \multirow{2}{*}{\cite{PVDIS:2014cmd}} & $-87.7$\cite{PVDIS:2014cmd} \\
scattering & $A_2 [10^{-6}]$ & $-160.8\pm7.1$ & & $-158.9$\cite{PVDIS:2014cmd}\\
 \cline{2-5}
 & \multirow{2}{*}{$g_{VA}^{eu}-g_{VA}^{ed}$} & $-0.042\pm0.057$ & \cite{Beise:2004py} & -0.0360\cite{ParticleDataGroup:2016lqr}\\
 & & $-0.12\pm0.074$ & \cite{Beise:2004py} & 0.0265\cite{ParticleDataGroup:2016lqr}\\
 \cline{2-5}
 & \multirow{2}{*}{$b_{\rm SPS}$ [GeV]$^{-2}$} & $-(1.47\pm0.42)\times10^{-4}$ & \cite{Argento:1982tq} & $-1.56\times10^{-4}$\cite{Argento:1982tq} \\
 & & $-(1.74\pm0.81)\times10^{-4}$ & \cite{Argento:1982tq} & $-1.57\times10^{-4}$\cite{Argento:1982tq} \\
\hline
{$\tau$} & $\mathcal{P}_\tau$ & $0.012\pm0.058$ & \multirow{2}{*}{\cite{VENUS:1997cjg}} & 0.028\cite{VENUS:1997cjg} \\
polarization  & $\mathcal{A}_{\mathcal{P}}$ & $0.029\pm0.057$ &  & 0.021\cite{VENUS:1997cjg} \\
\hline
{$\nu$ trident } & \multirow{2}{*}{${\sigma}/{\sigma^{\rm SM}}$} & \multirow{2}{*}{$0.82\pm0.28$} & \multirow{2}{*}{\cite{CHARM-II:1990dvf,CCFR:1991lpl,Altmannshofer:2014pba}} & \multirow{2}{*}{1} \\
{production} & & & &  \\
\hline
{$d_I\to u_J\ell\bar{\nu}_\ell(\gamma)$}& $\epsilon_{L,R,S,P,T}^{de_J}$ & \cite{Gonzalez-Alonso:2016etj} & \cite{Gonzalez-Alonso:2016etj} & 0 \\
\hline\hline
{$pp\to\ell^+\ell^-$}& $A_4$ & \cite{ATLAS:2018gqq} & \cite{ATLAS:2018gqq} & 0 \\
\hline
\end{tabular}
\end{adjustbox}
\end{table}
 
For the fit without \neffe, dubbed ``current fit'' in the following, we use the same input as summarized in \cite{deBlas:2022ofj} except that we do not include any data from future experiments. \yong{These are summarized in table\,\ref{tab:zwpoleobs} for electroweak precision observables, table\,\ref{tab:fpairobs} for light fermion pair production, and table\,\ref{tab:leobs} for the low-energy observables and the forward-backward asymmetries in Drell-Yan processes at the LHC. Definitions of these observables can be found in the corresponding references, which we do not repeat here.} When including \neffe, dubbed ``+CMB-S4'' for example when including the projection of \neff from CMB-S4, we use the following data/projections from Planck\,\cite{Aghanim:2018eyx}, CMB-S4\,\cite{Abazajian:2016yjj}, and CMB-HD\,\cite{Sehgal:2019ewc,CMB-HD:2022bsz}, respectively,
\eqal{
N_{\rm eff}^{\rm Planck} = & 2.99\pm0.34,\quad N_{\rm eff}^{\rm CMB-S4} = 2.99\pm0.027,\nb\\
&\, N_{\rm eff}^{\rm CMB-HD} = 2.99\pm0.014,\label{eq:neffexp}
}
assuming the central value of \neff stays the same as the current one from Planck.\footnote{The central value of \neff from CMB-S4/HD is unknown at this stage as the corresponding experiment is not performed yet. A different central value will only affect the best-fit points, which will not affect our conclusion in this work.}

\section{Results}\label{sec:results}
A global fit is then performed for the 61 parameters simultaneously\footnote{These 61 parameters are the same as those in \cite{deBlas:2022ofj}. Besides the inclusion of \neffe, the other difference from \cite{deBlas:2022ofj} is that we now fit to the real data in this work.} \yong{using the datasets summarized in Tables\,\ref{tab:zwpoleobs}, \ref{tab:fpairobs} and \ref{tab:leobs} and numbers of \neff in eq.\,\eqref{eq:neffexp}. In addition, whenever reported in the corresponding original references, we include the full correlations for such a global analysis. The} results are presented below in the flavor universal case in section\,\ref{sec:fucfit} and the flavor general case in section\,\ref{sec:fgcfit}.
\subsection{Flavor universal case}\label{sec:fucfit}
In this section, we present the global fit results with the inclusion of \neff from precision cosmology in the flavor universal case, and the results are shown in eq.\,\eqref{gfit:flavoruniv} {with the first number in each column the central value and the second number in parenthesis the corresponding 1$\sigma$ uncertainty}. The left column corresponds to the current fit, and the right column corresponds to the global fit with \neff from CMB-S4. Impact on the global fit from the current value of \neff from Planck is marginal compared with the current fit, we thus do not include them here. Similarly, the global fit results using \neff from CMB-HD are very close to those in the CMB-S4 case and are therefore not shown explicitly.

From eq.\,\eqref{gfit:flavoruniv}, we find that the overall impact of \neff on the SMEFT global fit is marginal even for CMB-S4/HD in the flavor universal case. The reason is that $c_{ll,l e, ee}$ can already be well constrained using the differential cross section of Bhabha scattering measured at LEP\,\cite{ALEPH:2005ab,ALEPH:2013dgf}. Note that these constraints will be surpassed by the MOLLER experiment\,\cite{MOLLER:2014iki} that is planned to start data collection in 2026\,\cite{mollerexp1,mollerexp2}, or future lepton colliders such as CLIC\,\cite{CLICPhysicsWorkingGroup:2004qvu,Robson:2018zje}, ILC\,\cite{ILC:2013jhg,ILCInternationalDevelopmentTeam:2022izu}, CEPC\,\cite{CEPCStudyGroup:2018ghi,CEPCPhysicsStudyGroup:2022uwl}, FCC-ee\,\cite{FCC:2018evy,Bernardi:2022hny}, or a muon collider\,\cite{Ankenbrandt:1999cta,Forslund:2022xjq,deBlas:2022aow}. For the electroweak vertices, they are constrained at or even below the per-mille level, thanks to the $Z/W$ pole runs of LEP. The helicity-conserving operators, i.e., $c_{ll}\to c_{ld}$, are generically constrained at the percent level, limited by the energy and luminosity of LEP. In contrast, the helicity-violating ones, i.e., $c_{lequ}\to c_{ledq}^{(3)}$, are also constrained at or below the per-mille level, due to (semi-)leptonic kaon decay from flavor studies\,\cite{Gonzalez-Alonso:2016etj}. In summary, we find the data consistent with the SM within $\sim2\sigma$.

\subsection{Flavor general case}\label{sec:fgcfit}
Global fit results in the most general flavor scenario are summarized in this section in eqs.\,(\ref{eq:gfit1}-\ref{eq:gfit3}). Again, the left column summarizes the results from the current fit, and the right column those by including \neff from CMB-S4. Results from the inclusion of \neff from the Planck are very close to the current fit results, we thus do not include them here. This is similarly true when comparing CMB-S4 with CMB-HD. Again, in the most general scenario, we find the improvement on SMEFT global fit from the inclusion of \neff is marginal. This can be understood from eq.\,\eqref{eq:neffsmeft}: Since those corrections from the operators are of $\mathcal{O}(10^{-3}\sim10^{-2})$ even when one takes $\mathcal{O}(1)$ values for the Wilson coefficients, these modifications can hardly be captured by future \neff experiments with $\sim1\%$ precision. Furthermore, since these Wilson coefficients are already stringently constraints to be around or even below $1\%$, it seems very challenging to use \neff to explore heavy new physics within the validity of the SMEFT given the current/planned precision on \neffe.

\begin{widetext}
\eqal{
\label{gfit:flavoruniv}
\begin{pmatrix}
 \\
\, \delta g_{Wl} \\ 
\, \delta g_{Z,L}^l \\ 
\, \delta g_{Z,R}^l  \\ 
\, \delta g_{Z,L}^u \\ 
\, \delta g_{Z,R}^u \\ 
\, \delta g_{Z,L}^d \\ 
\, \delta g_{Z,R}^d \\ 
\, c_{ll} \\
\, c_{le} \\
\, c_{ee} \\
\, c_{lq} \\
\, c_{lq}^{(3)} \\
\, c_{eu} \\
\, c_{ed} \\
\, c_{eq} \\
\, c_{lu} \\
\, c_{ld} \\
\, c_{lequ} \\
\, c_{lequ}^{(3)} \\
\, c_{ledq} \\
\, c_{ledq}^{(3)} \\
\end{pmatrix}
= 
\begin{pmatrix}
{\rm current\, fit} \\
-0.024(0.036) \\
-0.023(0.019) \\
-0.0091(0.021) \\
-0.14(0.17)\\
-0.16(0.34)\\
0.19(0.14)\\
1.5(0.73)\\
-0.29(0.43)\\
-0.17(0.13)\\
0.071(0.13)\\
0.15(0.31)\\
-1.9(0.89)\\
-0.67(0.52)\\
0.26(1.1)\\
3.8(1.9)\\
-0.57(2.2)\\
1.4(2.5)\\
2.7(4.0)\\
-0.078(0.075)\\
-0.024(0.19)\\
-0.078(0.075)\\
-0.25(7.1)\\
\end{pmatrix}\times 10^{-2}
\to
\begin{pmatrix}
\text{+CMB-S4} \\
-0.024(0.036) \\
-0.023(0.019) \\
-0.0091(0.021) \\
-0.14(0.17)\\
-0.16(0.34)\\
0.19(0.14)\\
1.5(0.73)\\
-0.29(0.43)\\
-0.17(0.13)\\
0.071(0.13)\\
0.15(0.31)\\
-1.9(0.89)\\
-0.67(0.52)\\
0.26(1.1)\\
3.8(1.9)\\
-0.57(2.2)\\
1.4(2.5)\\
2.7(4.0)\\
-0.078(0.075)\\
-0.024(0.19)\\
-0.078(0.075)\\
-0.25(7.1)\\
\end{pmatrix}\times 10^{-2} ,
}
\end{widetext}

\yong{Finally, we comment on the renormalization group evolution (RGE) of the SMEFT operators given the large energy gap between the SMEFT validity scale around 1\,TeV and that of the neutrino decoupling around 2\,MeV. In principle, this shall be done in multiple steps: One firstly runs the SMEFT operators from the TeV scale down to the weak scale, and then integrates out the heavy degrees of freedom of the SM, i.e., the top quark, the SU(2) gauge bosons, and the Higgs boson to obtain the low-energy effective filed theory at the weak scale. This same procedure is then repeated until the remaining light degrees of freedom are $e^\pm$ and the neutrinos, as appropriate for neutrino decoupling in the early Universe\,\cite{Jenkins:2013zja,Jenkins:2013wua,Alonso:2013hga,Jenkins:2017jig,Jenkins:2017dyc}. In practice, one notes that the running effects include those for the gauge couplings in eq.\,\eqref{eq:vertex} and those for the Wilson coefficients in eq.\,\eqref{eq:neffsmeft}, the later of which could again be interpreted as parameters defined at 1\,TeV for instance. For both of them, they are free of any strong interactions or any interference between electromagnetic/weak and strong interactions at the level of approximation we are considering in this work. As a consequence, the RGE effect is expected to be negligible, which we check explicitly with the help of {\tt Wilson}\,\cite{Aebischer:2018bkb} and find to be at the percent level. On the other hand, given the small corrections to \neff from these operators as seen in eq.\,\eqref{eq:neffsmeft} and the very stringent constraints on these Wilson coefficients from the global analysis as summarized in eqs.\,(\ref{eq:gfit1}-\ref{eq:gfit3}), the running effects need to be at least of about $50\gg g^2_{L,Y}\log({\rm 1TeV^2/2MeV^2})$ to become comparable with the uncertainty in \neff from CMB-S4 and even larger to have any impact on the global analysis. For these reasons, we claim that corrections from the RGE of these parameters can be safely neglected in this analysis.}

\section{\boldmath $Y_P$ and the neutron lifetime anomaly}\label{sec:ypneutron}
With the results presented in last section, one can then compute \neff and its uncertainty from the global fit. The uncertainty of \neff also affects the precision of $Y_P$, the primordial abundance of helium, through error propagation. In addition, the uncertainty of $Y_P$ also depends on that in the free neutron lifetime. Therefore, \neff from the global fit may provide, from a different angle, information on the free neutron lifetime anomaly. This section is devoted to this point from the global fit.
\subsection{Brief overview of the free neutron lifetime anomaly}
Free neutrons have a finite lifetime, they can decay weakly through $n\to p+e^-+\bar\nu_e$, with a lifetime $\tau_n$ of about 15 minutes\,\cite{ParticleDataGroup:2022pth}. Due to the flavor- and mass-eigenstate misalignment, the decay amplitude depends on the $V_{ud}$ element of the Cabibbo-Kobayashi-Maskawa (CKM) matrix\,\cite{Cabibbo:1963yz,Kobayashi:1973fv}. Therefore, a precision measurement of $\tau_n$ can play a precision role in terms of SM test through checking the unitarity of the CKM matrix. We refer the readers to the seminal paper by Sirlin\,\cite{Sirlin:1977sv} in radiative corrections to $\tau_n$ and \cite{Marciano:2005ec,Seng:2018yzq} for the recent progress.
\begin{widetext}
\eqal{
\label{eq:gfit1}
\begin{pmatrix}
 \\
\, \delta g_{W}^{e\nu} \\ 
\, \delta g_{W}^{\mu\nu} \\ 
\, \delta g_{W}^{\tau\nu} \\ 
\, \delta g_{Z,L}^{ee} \\ 
\, \delta g_{Z,L}^{\mu\mu} \\ 
\, \delta g_{Z,L}^{\tau\tau} \\ 
\, \delta g_{Z,R}^{ee} \\ 
\, \delta g_{Z,R}^{\mu\mu} \\ 
\, \delta g_{Z,R}^{\tau\tau} \\ 
\, \delta g_{Z,L}^{uu} \\ 
\, \delta g_{Z,L}^{cc} \\ 
\, \delta g_{Z,L}^{tt} \\ 
\, \delta g_{Z,R}^{uu} \\ 
\, \delta g_{Z,R}^{cc} \\ 
\, \delta g_{Z,L}^{dd} \\ 
\, \delta g_{Z,L}^{ss} \\ 
\, \delta g_{Z,L}^{bb} \\ 
\, \delta g_{Z,R}^{dd} \\ 
\, \delta g_{Z,R}^{ss} \\ 
\, \delta g_{Z,R}^{bb} \\ 
\, \delta g^{Wq_1}_R \\
\end{pmatrix}
= 
\begin{pmatrix}
{\rm current\, fit} \\
-1.0(0.64) \\
-1.4(0.59) \\
1.9(0.79)  \\
-0.023(0.027) \\
 0.012(0.12) \\
 0.019(0.059) \\
-0.031(0.027) \\
 0.0054(0.14) \\
 0.042(0.062) \\
-2.1(2.3) \\ 
-0.15(0.36) \\
0.13(4.1) \\
-1.1(2.8)\\
-0.35(0.53) \\
-2.9(3.4) \\
1.6(2.5) \\
0.33(0.17)\\
-3.0(12) \\
3.3(4.9) \\
2.3(0.87) \\
-1.3(1.7) \\
\end{pmatrix}\times 10^{-2} 
\to 
\begin{pmatrix}
\text{+CMB-S4} \\
-1.0(0.64) \\
-1.4(0.59) \\
1.9(0.79)  \\
-0.023(0.027) \\
 0.012(0.12) \\
 0.019(0.059) \\
-0.031(0.027) \\
 0.0054(0.14) \\
 0.042(0.062) \\
-2.1(2.3) \\ 
-0.15(0.36) \\
0.13(4.1) \\
-1.1(2.8)\\
-0.35(0.53) \\
-2.9(3.4) \\
1.6(2.5) \\
0.33(0.17)\\
-3.0(12) \\
3.3(4.9) \\
2.3(0.87) \\
-1.3(1.7) \\
\end{pmatrix}\times 10^{-2} ,
}
\eqal{
\label{eq:gfit2}
\begin{pmatrix}
  \\
\, [c_{ll}]_{1111} \\ 
\, [c_{le}]_{1111} \\     
\,  [c_{ee}]_{1111} \\    
\, [c_{ll}]_{1221} \\ 
\, [c_{ll}]_{1122} \\ 
\, [c_{le}]_{1122} \\ 
\, [c_{le}]_{2211} \\   
\, [c_{ee}]_{1122} \\    
\,  [c_{ll}]_{1331}  \\     
\, [c_{ll}]_{1133} \\ 
\, [c_{le}]_{1133} \\ 
\, [c_{le}]_{3311} \\
\,  [c_{ee}]_{1133} \\ 
\, [\hat c_{ll}]_{2222} \\ 
\, [c_{ll}]_{2332} \\
\end{pmatrix}
= 
\begin{pmatrix}
{\rm current\, fit} \\
 0.17(0.40) \\
-4.8(1.6) \\
 0.59(1.8) \\
 1.6(1.9) \\
-1.4(1.8) \\
4.3(2.4) \\
1.5(1.3) \\
-0.11(16) \\
-2.3(10) \\
1.7(10) \\
-1.3(17) \\
10(23) \\ 
3.0(2.3) \\
-1.5(2.9) \\
79(180) \\
\end{pmatrix}
\times 10^{-2}
\to
\begin{pmatrix}
\text{+CMB-S4} \\
 0.17(0.40) \\
-4.8(1.6) \\
 0.59(1.8) \\
 1.6(1.9) \\
-1.4(1.8) \\
4.3(2.4) \\
1.5(1.3) \\
-0.092(16) \\
-2.3(10) \\
1.7(10) \\
-1.3(17) \\
10(23) \\ 
3.0(2.3) \\
-1.5(2.9) \\
79(180) \\
\end{pmatrix}
\times 10^{-2},
}
\end{widetext}

\begin{widetext}
\eqal{
\label{eq:gfit3}
\begin{pmatrix}
  \\
\, [c_{lq}^{(3)}]_{1111}  \\  
\, [\hat c_{e q}]_{1111} \\
\, [\hat c_{l u}]_{1111} \\  
\, [\hat c_{l d}]_{1111}  \\  
\, [\hat c_{eu}]_{1111} \\
\, [\hat c_{ed}]_{1111} \\ 
\, [\hat c_{lq}^{(3)}]_{1122}  \\  
\, [c_{lu}]_{1122}  \\
\, [\hat c_{ld}]_{1122} \\ 
\, [c_{eq}]_{1122} \\ 
\, [c_{eu}]_{1122}  \\  
\, [\hat c_{ed}]_{1122} \\  
\, [\hat c_{lq}^{(3)}]_{1133} \\
\, [c_{ld}]_{1133} \\
\, [c_{eq}]_{1133} \\
\, [c_{ed}]_{1133} \\
\, [c_{lq}^{(3)}]_{2211} \\
\, [c_{lq}]_{2211} \\
\, [c_{lu}]_{2211} \\
\, [c_{ld}]_{2211} \\
\, [\hat c_{eq}]_{2211} \\
\, [c_{lequ}]_{1111} \\
\, [c_{ledq}]_{1111} \\
\, [c_{lequ}^{(3)}]_{1111} \\
\, \epsilon_P^{d\mu}[{\rm 2\,GeV}] \\
\end{pmatrix}
= 
\begin{pmatrix}
{\rm current\, fit} \\
0.91(0.40) \\
-0.22(0.22)\\
0.51(8.8) \\
4.3(18) \\
-4.2(9.4) \\
-9.4(17) \\
-50(31) \\
1.9(7.6) \\
-260(128) \\
-18(25) \\
-71(45) \\
221(134) \\
-3.7(7.2) \\
1.3(10) \\
-1.9(5.0) \\
6.2(18) \\
-0.45(3.6) \\
-2.0(5.9) \\
9.8(8.1) \\
12(26) \\
0.77(41) \\
-0.082(0.076) \\
-0.081(0.076) \\
-0.024(0.19) \\
0.021(0.14) \\
\end{pmatrix}
\times 10^{-2}
\to
\begin{pmatrix}
\text{+CMB-S4} \\
0.91(0.40) \\
-0.22(0.22)\\
0.51(8.8) \\
4.3(18) \\
-4.2(9.4) \\
-9.4(17) \\
-50(31) \\
1.9(7.6) \\
-260(128) \\
-18(25) \\
-71(45) \\
221(134) \\
-3.7(7.2) \\
1.3(10) \\
-1.9(5.0) \\
6.2(18) \\
-0.45(3.6) \\
-2.0(5.9) \\
9.8(8.1) \\
12(26) \\
0.76(40) \\
-0.082(0.076) \\
-0.081(0.076) \\
-0.024(0.19) \\
0.021(0.14) \\
\end{pmatrix}
\times 10^{-2},
}
\end{widetext}

The leading $\tau_n$ measurements come from two different and independent groups using the beam and the bottle methods, respectively. In the former case, it is the decay product $p$ of free neutrons that is measured, and in the latter case, it is the number of surviving free neutrons that is counted. Currently, the most precise result in the beam case is obtained by Sussex-ILL-NIST, with
\eqal{
\tau_n = (887.7\pm 1.2\,{\rm [stat.]} \pm 1.9{\rm [syst.]})\,s\quad\text{\cite{Yue:2013qrc}}.
}
Note that the current uncertainty in the beam case is still above 1 second. In the bottle case, this uncertainty has recently been reduced below 1 second in \cite{Pattie:2017vsj,UCNt:2021pcg} using ultra cold neutrons to reduce wall loss, and the latest result from UCN$\tau$ is
\eqal{
\tau_n = (877.75\pm 0.28\,{\rm [stat.]}^{+0.22}_{-0.16}\,{\rm [syst.]})\,s\quad\text{\cite{UCNt:2021pcg}}.
}
These two most precise results differ by about 10 seconds in their central values, which is beyond $3\sigma$ and known as the neutron lifetime anomaly. For a recent review on this topic, see \cite{Wietfeldt:2011suo}. This anomaly could possibly be accounted for by exotic neutron decay(s), though the corresponding decay branching ratio is highly constrained\,\cite{Fornal:2018eol,Czarnecki:2018okw,Dubbers:2018kgh}.

As mentioned at the beginning of this section, the uncertainty in $\tau_n$ enters that in $Y_P$ in the Standard Big-Bang Nucleosynthesis (SBBN) calculation. The same is true in \neffe, whose absolute theoretical uncertainty is currently at the $10^{-4}$ level as discussed earlier. In this work, we ask ourselves the following question: Given the global fit results presented in last section, what is their impact on $Y_P$ and what light it can shed on the neutron lifetime anomaly? We will try to answer this question in the next subsection.

\subsection{From SMEFT global fit to the primordial helium abundance}
\begin{figure}[!htb]
\begin{center}
\includegraphics[width=\columnwidth]{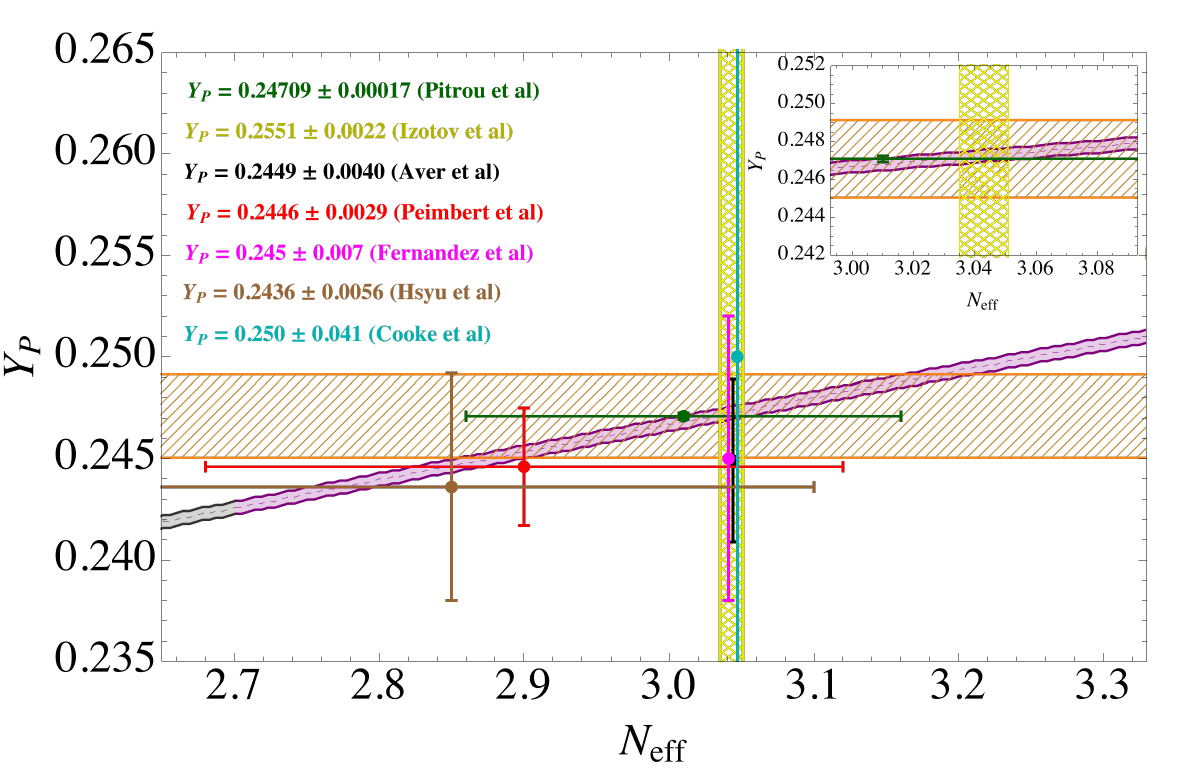}
\caption{The plot for the predicted primordial Helium abundance $Y_P$ as a function of \neffe. The standard model primordial value $Y_P=0.24709\pm0.00017$\,\cite{Pitrou:2018cgg} at 95\% confidence level (CL) is indicated by the green cross. The predicted 95\% CL value of $Y_P$ {with \neff in the Planck range} is given by the diagonal band, where the dashed line represents its central value. The gray band in the lower left corner is disfavored by BBN. \neff from the global fit is shown by the vertical hatched region in yellow. The inset at the upper right corner represents the enlarged intersection between the yellow hatched region and the diagonal band in purple. The other crosses are experimental measurements of $Y_P$ from different groups as indicated by the legends (see the text for more details).}\label{fig:neff2yp}
\end{center}
\end{figure}

With the $1\sigma$ bounds for the electroweak vertices and the 4-fermion operators in eqs.\,(\ref{eq:gfit1}-\ref{eq:gfit2}), we randomly varying the dependent Wilson coefficients in eq.\,\eqref{eq:neffsmeft} inside the 13-dimensional $1\sigma$ sphere with $10^9$ points to fit \neff and obtain
\eqal{
N_{\rm eff} = 3.043 \pm 0.008,\label{eq:nefffit}
}
which is consistent with the current measurement from Planck\,\cite{Aghanim:2018eyx} and the SM prediction\,\cite{Akita:2020szl,Froustey:2020mcq} within $1\sigma$. Note also that the error of \neff in eq.\,\eqref{eq:nefffit} is of the same order as its current theoretical uncertainty. Since the error would be further reduced, for example, at future lepton colliders\,\cite{deBlas:2022ofj}, the theoretical uncertainty will become important and shall be included for the study at these next generation experiments. We then make use of {\tt AlterBBN}\,\cite{Arbey:2018zfh} to predict the primordial abundance of helium $Y_P$ with \neff being the only free parameter but varying in the Planck range in eq.\,\eqref{eq:neffexp} and by fixing the neutron lifetime $\tau_n$ and its uncertainty $\Delta \tau_n$ from\,\cite{UCNt:2021pcg}. The result is shown in Fig.\,\ref{fig:neff2yp} by the diagonal band,\footnote{If instead the beam values of $\tau_n$ and $\Delta \tau_n$ are used\,\cite{Yue:2013qrc}, the width of this band will increase and the other parts of this plot will remain untouched.} whose half-width corresponds to the uncertainty in $Y_P$. The gray part in the lower-left corner is disfavored by BBN, ignoring the correlation of $^4{\rm He}$ with the other light elements. The predicted $Y_P=0.24709\pm0.00017$\,\cite{Pitrou:2018cgg} from the SBBN is indicated by the green cross, whose absolute error is given by the vertical bar and is invisible in the plot. The fitted \neff in eq.\,\eqref{eq:nefffit} is then shown by the vertical hatched region in yellow, whose intersection with the diagonal band in purple is shown by the inset in the upper right corner. There, one can clearly see that \neff from the global fit significantly narrows down the space on the $Y_P$-\neff plane, corresponding to a significant improvement compared with the diagonal band in the Planck scenario.

Experimentally, $Y_P$ can be measured from metal-poor galaxies by observing the $^4{\rm He}$ and hydrogen emission lines as a function of the galaxy metallicity and extrapolating back to the zero metallicity point for the estimation. In \cite{Izotov:2014fga,Aver:2015iza,Peimbert:2016bdg,Fernandez:2018xyz,Hsyu:2020uqb}, they obtained, respectively,
\eqal{
Y_P  &\, = 0.2551 \pm 0.0022,\nb\\
Y_P  &\, = 0.2449 \pm 0.0040, \nb\\
Y_P  &\, = 0.2446 \pm 0.0029, \nb\\
Y_P  &\, = 0.245 \pm 0.007, \nb\\
Y_P &\, = 0.2436_{-0.0040}^{+0.0039}.
}
In addition, from the intergalactic medium, \cite{Cooke:2018qzw} recently found
\eqal{
Y_P=0.250_{-0.025}^{+0.033}.
}
Note that the error in the latter case is much larger than that from metal-poor galaxies. All these experimental results are shown in order in Fig.\,\ref{fig:neff2yp} as indicated by the legend, with the outlier from\,\cite{Izotov:2014fga} beyond the scope of this plot. From this figure, we find all these experimental results also consistent with \neff from the global fit but with large uncertainties.

Recall that the uncertainty in $Y_P$ is directly related to that in neutron lifetime\,\cite{Pitrou:2018cgg}, we show in Fig.\,\ref{fig:neff2yp} the uncertainty band of $Y_P$ in hatched orange to show explicitly the region \textit{insufficient} to resolve the $\sim10\,s$ neutron lifetime anomaly. We stress that, to get this hatched region, we assume the central value of $Y_P$ is aligned with the SBBN prediction in\,\cite{Pitrou:2018cgg}, and use the central value of $\tau_n$ from\,\cite{Pattie:2017vsj,UCNt:2021pcg} as its uncertainty is small below 1\,s.\footnote{We do not use the world average in \cite{ParticleDataGroup:2022pth} here as we want to account for the difference in $\tau_n$ from different experiments. In addition, \cite{Aver:2015iza,Fernandez:2018xyz,Cooke:2018qzw} did not report any results on \neffe, we thus use its SM value when plotting their data in Fig.\,\ref{fig:neff2yp}. A spurious shift of 0.003 to \neff is then applied to separate these data, in black, magenta, and cyan respectively, for a better visibility.} Clearly, the SBBN result precludes its possibility in accounting for the difference between the beam and the bottle results, as is well-known. On the other hand, an improvement in extracting $Y_P$ from astrophysical observation, reducing the uncertainty in $Y_P$ to be within the orange band to be more quantitative, would deepen our understanding on the connection between SBBN and the neutron lifetime anomaly.\footnote{See, for example, \cite{1982AZh5915P,1983AZh32326P,1984AZh86970P,1984AZh27982P,1985AZh23638P,1988SvA32127D,1989SvA92338D,1995SvA13338D,1990SvA96103D,1991SvA7378D,1994SvA69097D,Khlopov:2013ava,Deng:2023twb,AristizabalSierra:2023bah}.} Here we stress that, at this point, the results in Fig.\,\ref{fig:neff2yp} are not obtained from the SMEFT global fit except the vertical hatched region in yellow. A full global analysis including all the light elements in the combined framework of SMEFT and SBBN would be desirable, but is beyond the scope of this work since it is more involved and will therefore be investigated in a future work.

\section{Conclusions}\label{sec:conclu}
In light of the percent level precision target of \neff from future cosmological probes like CMB-S4, we think it is now proper to combine the cosmological data together with the collider and the low-energy data sets available to perform a global fit of the SMEFT in searching for new physics indirectly. In this work, by utilizing the \neff data from Planck, as well as its projections for CMB-S4 and CMB-HD, we perform a global fit of the SMEFT without any flavor assumption in the electroweak vertex and (semi-)leptonic 4-fermion sectors. From this work, we find that:
\begin{enumerate}
\item In both the flavor universal and the most general flavor scenarios, the inclusion of \neff marginally improves the SMEFT global fit due to very small corrections to \neff from the SMEFT operators as seen in eq.\,\eqref{eq:neffsmeft}. The agreement on \neff between the SM prediction and the SMEFT global fit also suggests that the current/planned experimental precision of \neff is very challenging in studying heavy new physics model independently within the SMEFT, rooted in already very stringent constraints on these operators from colliders and low-energy experiments. On the other hand, thanks to these very strong constraints, the $1\sigma$ bound on \neff from the global fit is also significantly reduced and becomes comparable to its current theoretical uncertainty as seen in eq.\,\eqref{eq:nefffit}. Since this $1\sigma$ bound would be further reduced at future lepton colliders for example, the theoretical uncertainty in \neff will become important at that stage.
\item \neff from the global fit significantly narrows down the parameter space on the $Y_P$-\neff plane compared with that using \neff from Planck, as clearly seen in Fig.\,\ref{fig:neff2yp}. Therefore, improving the experimental precision for both \neff and $Y_P$ could add important inputs for a precision test of the SM and the SBBN.
\item By comparing $Y_P$ from astrophysical experiments and the global fitted result of \neffe, we find that improving the experimental precision on $Y_P$ from metal-poor galaxies will play an important role in our understanding on the neutron lifetime anomaly within the SBBN, as shown in Fig.\,\ref{fig:neff2yp}.
\end{enumerate}

In the flavor universal case, we commented that the precision in $[\mathcal{O}_{\ell\ell,\ell e, ee}]$ was currently limited by Bhabha scattering at LEP. Looking ahead into the near future, the MOLLER experiment at the Jefferson Laboratory, planned to start data collection in early 2026, will measure the weak mixing angle at a low momentum transfer at the per-mille level. We expect the global fit in this sector to be improved by at least a factor of a few in both the flavor universal and the flavor general cases. Furthermore, in the current work, we only consider $Y_P$ {by using the global fitted result of \neff} and leaving out the other light elements from the consideration of their current uncertainties. It would be desirable to include the full relevant SMEFT operators during BBN for studying \neffe, the primordial abundances of light elements, and its interplay with the neutron lifetime. This is left for a future work.

\acknowledgments{The author would like to thank Jorge de Blas, Christophe Grojean, Jiayin Gu, Victor Miralles, Michael E. Peskin, Junping Tian, Marcel Vos and Eleni Vryonidou for working on SMEFT global fit for Snowmass 2021\,\cite{deBlas:2022ofj}. This project was funded by China Postdoctoral Science Foundation under grant number 2023M732255, the Postdoctoral Fellowship Program of CPSF under number GZC20231613. YD also acknowledges support from the Shanghai Super Postdoc Incentive Plan, and the T.D. Lee Postdoctoral Fellowship at the Tsung-Dao Lee Institute, Shanghai Jiao Tong University.}

\appendix

\section{Analytical expression for $\delta\rho_\nu^{\rm tot}/\delta t$ in the SMEFT}\label{app:drhodt}
Similar to eq.\,\eqref{eq:neffdndt}, the total neutrino energy density changing rate, ${\delta\rho_\nu^{\rm tot}}/{\delta t} \equiv \sum\limits_{\alpha=e}^\tau {\delta\rho_{\nu_\alpha}}/{\delta t}$ including anti-neutrinos, is given here explicitly:
\begin{widetext}
\eqal{
\frac{\delta\rho_\nu^{\rm tot}}{\delta t} =&\,  \frac{8G_F^2}{\pi^5}\left[ 4 \left(8 s_W^4+4 s_W^2+1\right) \left(T_{\gamma}^9-T_{\nu_e}^9\right)+8 \left(8 s_W^4-4 s_W^2+1\right) \left(T_{\gamma}^9-T_{\nu_\mu}^9\right)\right.\nb\\
&\left.\qquad\quad +7 T_{\gamma}^4 \left(\left(8 s_W^4+4 s_W^2+1\right) T_{\nu_e}^4 (T_{\gamma}-T_{\nu_e})+2 \left(8 s_W^4-4 s_W^2+1\right) T_{\nu_\mu}^4 (T_{\gamma}-T_{\nu_\mu})\right) \right. \nb\\
&\left.\qquad\quad - 4 [c_{le}]_{1111} s_W^2 \left(4 \left(T_\gamma^9-T_{\nu_e}^9\right)+7 T_\gamma^4 T_{\nu_e}^4 (T_\gamma-T_{\nu_e})\right) \right. \nb\\
&\left.\qquad\quad - 2 [c_{ll}]_{1111} \left(2 s_W^2+1\right) \left(4 \left(T_\gamma^9-T_{\nu_e}^9\right)+7 T_\gamma^4 T_{\nu_e}^4 (T_\gamma-T_{\nu_e})\right) \right. \nb\\
&\left.\qquad\quad + 4 \delta g_{W}^{e\nu} \left(8 s_W^4+4 s_W^2+1\right) \left(4 T_\gamma^9+7 T_\gamma^5 T_{\nu_e}^4-7 T_\gamma^4 T_{\nu_e}^5-4 T_{\nu_e}^9\right) \right. \nb\\
&\left.\qquad\quad + 8 \delta g_{Z,L}^{ee} \left( 4T_\gamma^9 (4s_W^4 + 3s_W^2-1) - 4 T_{\nu_e}^9 (4s_W^4+s_W^2) + (4-8s_W^2)T_{\nu_\mu}^9\right.\right.\nb\\
&\left.\left.\qquad\qquad\qquad\qquad + \left(28s_W^4+7s_W^2\right) \left( T_\gamma^5 T_{\nu_e}^4 - T_\gamma^4 T_{\nu_e}^5 \right)  + \left(7-14s_W^2\right) \left( T_\gamma^4 T_{\nu_e}^5 - T_\gamma^5 T_{\nu_e}^4 \right) \right) \right. \nb\\
&\left.\qquad\quad + 8 \delta g_{Z,R}^{ee} s_W^2 \left(12 T_\gamma^9+7 T_\gamma^5 \left(T_{\nu_e}^4+2 T_{\nu_\mu}^4\right)-7 T_\gamma^4 \left(T_{\nu_e}^5+2 T_{\nu_\mu}^5\right)-4 \left(T_{\nu_e}^9+2 T_{\nu_\mu}^9\right)\right) \right. \nb\\
&\left.\qquad\quad - 4 \left( [c_{le}]_{2211} + [c_{le}]_{3311}  \right) s_W^2 \left(4 T_\gamma^9+7 T_\gamma^5 T_{\nu_\mu}^4-7 T_\gamma^4 T_{\nu_\mu}^5-4 T_{\nu_\mu}^9\right) \right. \nb\\
&\left.\qquad\quad + 2 \left( [c_{ll}]_{1122} + [c_{ll}]_{1133} \right) \left(1-2 s_W^2\right) \left(4 T_\gamma^9+7 T_\gamma^5 T_{\nu_\mu}^4-7 T_\gamma^4 T_{\nu_\mu}^5-4 T_{\nu_\mu}^9\right) \right. \nb\\
&\left.\qquad\quad + 4 \left( \delta g_{W}^{\mu\nu} + \delta g_{W}^{\tau\nu} + \delta g_{Z,L}^{\mu\mu} + \delta g_{Z,L}^{\tau\tau} \right) \left(8 s_W^4-4 s_W^2+1\right) \right.\nb\\
&\left.\qquad\qquad \times \left(4 T_\gamma^9+7 T_\gamma^5 T_{\nu_\mu}^4-7 T_\gamma^4 T_{\nu_\mu}^5-4 T_{\nu_\mu}^9\right)  \right],\label{eq:rhotot}
}
\end{widetext}
where small corrections from non-vanishing $m_e$ and Fermi-Dirac statistics are also ignored, and we fix $T_{\nu_\mu}=T_{\nu_\tau}$ and $\mu_{\nu_\alpha}=0$ for simplicity. Clearly, our argument for $\delta n_\nu^{\rm tot}/\delta t$ below eq.\,\eqref{eq:neffdndt} applies similarly for ${\delta\rho_\nu^{\rm tot}}/{\delta t}$. For example, since $s_W^2\approx 1/4$, the $\delta g_{Z,L}^{ee}$ term also becomes proportional to an overall factor $4 s_W^4 + 3 s_W^2 -1 \approx 0$ in the $T_{\nu_e}\approx T_{\nu_\mu} = T_{\nu_\tau}$ limit, explaining why the contribution to \neff from $\delta g_{Z,L}^{ee}$ is relatively small in eq.\,\eqref{eq:neffsmeft}.

\bibliography{ref,SMEFT21}

\end{document}